\documentclass[showpacs,preprintnumbers,amsmath,amssymb]{revtex4}

\usepackage{graphicx}
\usepackage{dcolumn}
\usepackage{bm}
\usepackage{epsf}

\newcommand{\beq}{\begin{equation}}
\newcommand{\eeq}{\end{equation}}
\newcommand\beqa{\begin{eqnarray}}
\newcommand\eeqa{\end{eqnarray}}
\newcommand\bea{\begin{array}}
\newcommand\eea{\end{array}}
\newcommand{\nn}{\nonumber}
\newcommand{\neqa}{\nonumber\end{eqnarray}}
\newcommand{\la}{\label}

\newcommand{\eq}[1]{eq.(\ref{#1})}
\newcommand{\eqs}[2]{eqs.(\ref{#1},\ref{#2})}
\newcommand{\Eq}[1]{Eq.(\ref{#1})}
\newcommand{\Eqs}[2]{Eqs.(\ref{#1},\ref{#2})}
\newcommand{\ur}[1]{(\ref{#1})}
\newcommand{\urs}[2]{(\ref{#1},\ref{#2})}

\newcommand{\Tr}{{\rm Tr}}
\newcommand{\cv}{{\cal V}}
\newcommand{\Det}{{\rm Det}}
\newcommand{\half}{\frac{1}{2}}
\renewcommand{\d}{\partial}
\renewcommand{\O}{{\cal O}}
\newcommand{\Mfunction}[1]{#1}
\newcommand{\<}{{\langle}}
\renewcommand{\>}{{\rangle}}
\newcommand{\D}{r_{12}}

\newcommand{\cA}{{\cal A}}
\newcommand{\cB}{{\cal B}}
\newcommand{\bv}{\overline{{\rm v}}}
\newcommand{\Mphi}{\Phi}
\renewcommand{\v}{{\rm v}}

\newcommand{\re}{\relax{\rm I\kern-.18em R}}

\newcommand{\sh}{{\rm sh_{\half}}}
\newcommand{\ch}{{\rm ch_{\half}}}
\newcommand{\bsh}{\overline{\sh}}
\newcommand{\bch}{\overline{\ch}}
\newcommand{\bshd}{\overline{\mathrm{sh}}\,}
\newcommand{\bchd}{\overline{\mathrm{ch}}\,}
\newcommand{\shd}{\mathrm{sh}\,}
\newcommand{\chd}{\mathrm{ch}\,}

\def\su2{{SU(2)}}
\def\cM{{\cal{M}}}
\def\tr{{\rm tr}}
\def\m{{\rm m}}
\def\s{{\rm s}}
\def\r{{\rm r}}

\begin{document}

\title{Quantum weights of dyons and of instantons with non-trivial holonomy}
\author{Dmitri Diakonov$^{a,b}$}
\author{Nikolay Gromov$^c$}
\author{Victor Petrov$^b$}
\author{Sergey Slizovskiy$^c$}
\vskip 0.3true cm

\affiliation{ $^a$NORDITA,  Blegdamsvej 17, DK-2100 Copenhagen,
Denmark\\
$^b$St. Petersburg NPI, Gatchina, 188 300, St. Petersburg, Russia\\
$^c$St. Petersburg State University, Faculty of Physics,
Ulianovskaya 1, 198 904, St. Petersburg, Russia. }

\date{April 9, 2004}

\begin{abstract}
We calculate exactly functional determinants for quantum
oscillations about periodic instantons with non-trivial value of
the Polyakov line at spatial infinity. Hence, we find the weight 
or the probability with which calorons 
with non-trivial holonomy occur in the Yang--Mills partition function. 
The weight depends on the value of the holonomy, the temperature, 
$\Lambda_{\rm QCD}$, and the separation between the BPS monopoles
(or dyons) which constitute the periodic instanton. At large
separation between constituent dyons, the quantum measure factorizes
into a product of individual dyon measures, times a definite
interaction energy. We present an argument that at temperatures below a 
critical one related to $\Lambda_{\rm QCD}$, trivial holonomy is
unstable, and that calorons ``ionize'' into separate dyons. 
\end{abstract}



\maketitle

\section{Motivation and the main result}

There are two known generalizations of the standard self-dual instantons 
to non-zero temperatures. One is the periodic instanton of Harrington and 
Shepard \cite{HS} studied in detail by Gross, Pisarski and Yaffe ~\cite{GPY}. 
These periodic instantons, also called {\em calorons}, are said to have 
trivial holonomy at spatial infinity. It means that the Polyakov line
\beq
L= \left.{\rm P}\,\exp\left(\int_0^{1/T}\!dt\,A_4\right)\right|_{|\vec x|\to\infty}
\la{Pol1}\eeq
assumes values belonging to the group center $Z(N)$ for the $SU(N)$
gauge group
\footnote{We use anti-Hermitian fields: $A_\mu= -i t^a A_\mu^a$,
$\Tr(t^at^b) =  \half\delta^{ab}$.}.
The vacuum made of those instantons has been investigated,
using the variational principle, in ref.~\cite{DMir}. \\

The other generalization has been constructed a few years ago by Kraan and
van Baal~\cite{KvB} and Lee and Lu~\cite{LL}; it has been named
{\em caloron with non-trivial holonomy} as the Polyakov line for this 
configuration does not belong to the group center. We shall call it for short the
KvBLL caloron. It is also a periodic self-dual solution of the Yang--Mills equations
of motion with an integer topological charge. In the limiting case
when the KvBLL caloron is characterized by trivial holonomy, it is reduced
to the Harrington--Shepard caloron. The fascinating feature of the KvBLL
construction is that a caloron with a unit topological charge can be viewed
as ``made of'' $N$ Bogomolnyi--Prasad--Sommerfeld (BPS) monopoles or
dyons~\cite{Bog,PS}. \\

Dyons are self-dual solutions of the Yang--Mills equations of motion with static
({\it i.e.} time-independent) action density, which have both the magnetic and
electric field at infinity decaying as $1/r^2$. Therefore these objects
carry both electric and magnetic charges (prompting their name).
In the $3\!+\!1$-dimensional $SU(2)$ gauge theory there are in fact two
types of self-dual dyons~\cite{LY}: $M$ and $L$ with (electric,  magnetic)
charges $(+,+)$ and $(-,-)$, and two types of anti-self-dual dyons
$\overline{M}$ and $\overline{L}$ with charges $(+,-)$ and $(-,+)$, respectively.
Their explicit fields can be found {\it e.g.} in ref.~\cite{DPSUSY}.
In the $SU(N)$ theory there are $2N$ different dyons~\cite{LY,DHKM}:
$M_1,M_2,...M_{N-1}$ ones with charges counted with respect to $N-1$ Cartan
generators and one $L$ dyon with charges compensating those of $M_1...M_{N-1}$
to zero, and their anti-self-dual counterparts. \\

Speaking of dyons one implies that the Euclidean space-time is compactified in
the `time' direction whose inverse circumference is temperature $T$, with
the usual periodic boundary conditions for boson fields. However,
the temperature may go to zero, in which case the $4d$ Euclidean invariance
is restored. \\

\begin{figure}[t]
\centerline{
\epsfxsize=0.6\textwidth
\epsfbox{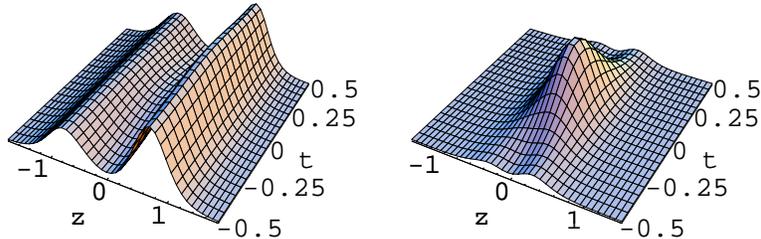}}
\caption{The action density of the KvBLL caloron as function
of $z,t$ at fixed $x=y=0$, with the asymptotic value of $A_4$ at spatial
infinity $\v=0.9\pi T,\; \bv=1.1\pi T$. It is periodic in $t$  direction.
At large dyon separation the density becomes static (left, $\D=1.5/T$).
As the separation decreases the action density becomes more like a $4d$
lump (right, $\D=0.6/T$). In both plots the L,M dyons are centered at
$z_{\rm L}=-\v\,\D/2\pi T,\;z_{\rm M}=\bv\,\D/2\pi T,\;x_{\rm L,M}=y_{\rm L,M}=0$.
The axes are in units of temperature T.\label{fig:adp1}}
\end{figure}

Dyons' essence is that the $A_4$ component of the dyon field tends
to a constant value at spatial infinity. This constant $A_4$
can be eliminated by a time-dependent gauge transformation. However then
the fields violate the periodic boundary conditions, unless $A_4$ has
quantized values corresponding to trivial holonomy, {\it i.e.} unless
the Polyakov line belongs to the group center. Therefore, in a general
case one implies that dyons have a non-zero value of $A_4$ at spatial
infinity and a non-trivial holonomy. \\

The KvBLL caloron of the $SU(2)$ gauge group (to which we restrict ourselves
in this paper) with a unit topological charge is ``made of" one $L$ and one
$M$ dyon, with total zero electric and magnetic charges. Although
the action density of isolated $L$ and $M$ dyons does not depend on time,
their combination in the KvBLL solution is generally non-static: the
$L,M$ ``constituents'' show up not as $3d$ but rather as $4d$ lumps, see
Fig.~1. When the temperature goes to zero while the separation between
dyons remain fixed, these lumps merge, and the KvBLL caloron
is reduced to the usual Belavin--Polyakov--Schwarz--Tyutin  instanton~\cite{BPST}
(as is the standard Harrington--Shepard caloron), {\em plus} corrections
of the order of $T$. However, the holonomy remains fixed and non-trivial
at spatial infinity. \\

There is a strong argument against the presence of either dyons or
KvBLL calorons in the Yang--Mills partition function at nonzero
temperatures~\cite{GPY}. The point is, the 1-loop effective action obtained from
integrating out fast varying fields where one keeps all powers of $A_4$
but expands in (covariant) derivatives of $A_4$ has the form \cite{DO}
\beqa
\nn
S^{\rm 1-loop}&= &\int\!d^4x\,\left[P(A_4)+E^2f_E(A_4)+B^2f_B(A_4)
+ {\rm higher\; derivative\; terms}\right],\\
\la{Tpot}
P(A_4)&=&\left.\frac{1}{3T(2\pi)^2}\v^2(2\pi T-\v)^2\right|_{{\rm  mod}\; 2\pi T},\qquad
\v=\sqrt{A_4^aA_4^a}\qquad[{\rm for\;the}\;SU(2)\;{\rm group}]
\eeqa
where the perturbative potential energy term $P(A_4)$ has been known for  a long
time \cite{GPY,NW}, see Fig.~2.  As follows from \eq{Pol1} the trace of  the
Polyakov line is related to $\v$ as
\beq
\half\,\Tr\,L= \cos\frac{\v}{2T}.
\la{TrL}\eeq
The zeros of the potential energy correspond to $\half\Tr\,L=\pm 1$, {\it i.e.}
to the trivial holonomy. If a dyon has $\v\neq 2\pi T n$ at spatial infinity
the potential energy is positive-definite and proportional to the $3d$
volume. Therefore, dyons and KvBLL calorons with non-trivial holonomy seem
to be strictly forbidden: quantum fluctuations about them have an unacceptably
large action. \\

Meanwhile, precisely these objects determine the physics of the
supersymmetric YM theory where in addition to gluons there are gluinos,
{\it i.e.} Majorana (or Weyl) fermions in the adjoint representation. Because
of supersymmetry, the boson and fermion determinants about $L,M$ dyons cancel exactly,
so that the perturbative potential energy \ur{Tpot} is identically zero for
all temperatures, actually in all loops. Therefore, in the  supersymmetric theory
dyons are openly allowed. [To be more precise, the cancellation occurs when
periodic conditions for gluinos are imposed, so it is the compactification
in one (time) direction that is implied, rather than physical temperature
which requires antiperiodic fermions.] Moreover, it turns out
\cite{DHKM} that dyons generate a non-perturbative potential having a
minimum at $\v=\pi T$, {\em i.e.} where the perturbative potential would
have the maximum. This value of $A_4$ corresponds to the holonomy
$\Tr\,L=0$ at spatial infinity, which is the ``most non-trivial'';
as a matter of fact $<\!\Tr\,L\!>=0$ is one of the confinement's requirements.
In the supersymmetric YM theory configurations having $\Tr\,L=0$ at
infinity are not only allowed but dynamically preferred as compared to
those with $\half\Tr\,L=\pm 1$. In non-supersymmetric theory it looks
as if it is the opposite. \\

Nevertheless, it has been argued in ref. \cite{Dobzor} that the  perturbative
potential energy \ur{Tpot} which forbids individual dyons in the pure YM
theory might be overruled by non-perturbative contributions of an {\em ensemble}
of dyons. For fixed dyon density, their number is proportional to the
$3d$ volume and hence the non-perturbative dyon-induced potential as  function
of the holonomy (or of $A_4$ at spatial infinity) is also proportional  to the volume.
It may be that at temperatures below some critical one the  non-perturbative potential
wins over the perturbative one so that the system prefers  $<\!\Tr\,L\!>=0$.
This scenario could then serve as a microscopic mechanism of the confinement-deconfinement
phase transition \cite{Dobzor}. It should be noted that the KvBLL calorons and
dyons seem to be observed in lattice simulations below the phase transition
temperature \cite{Brower,IMMPSV,Gatt}. \\

To study this possible scenario quantitatively, one first needs to find out
the quantum weight of dyons or the probability with which they appear
in the Yang--Mills partition function. Unfortunately, the single-dyon measure 
is not well defined: it is too badly divergent in the infrared region 
owing to the weak (Coulomb-like) decrease of the fields. What makes sense 
and is finite, is the quantum determinant for small oscillations about
the KvBLL caloron which possesses zero net electric and magnetic charges.
To find this determinant is the primary objective of this study. 
The KvBLL measure depends on the asymptotic value of $A_4$ (or on the holonomy 
through \eq{TrL}), on the temperature $T$, on $\Lambda$, the scale parameter 
obtained through the renormalization of the charge, and on the dyon separation $\D$. 
At large separations between constituent $L,M$ dyons of the caloron, one 
gets their weights and their interaction. \\

The problem of computing the effect of quantum fluctuations about a caloron with
non-trivial holonomy is of the same kind as that for ordinary instantons (solved
by 't Hooft~\cite{tHooft}) and for the standard Harrington--Shepard caloron (solved
by Gross, Pisarski and Yaffe~\cite{GPY}) being, however, technically much more difficult.
The zero-temperature instanton is $O(4)$ symmetric, and the  Harrington--Shepard
caloron is $O(3)$ symmetric, which helps. The KvBLL caloron has no such symmetry
as obvious from Fig.~1. Nevertheless, we have managed to find the small-oscillation
determinant {\em exactly}. It becomes possible because we are able to construct the
exact propagator of spin-0, isospin-1 field in the KvBLL background, which by itself
is some achievement. \\

As it is well known~\cite{tHooft,GPY}, the calculation of the quantum weight
of a Euclidean pseudoparticle consists of three steps: i) calculation of the
metric of the moduli space or, in other words, computing the Jacobian composed
of zero modes, needed to write down the pseudoparticle measure in terms of its
collective coordinates, ii) calculation of the functional determinant for
non-zero modes of small fluctuations about a pseudoparticle, iii) calculation of
the ghost determinant resulting from background gauge fixing in the previous step.
Problem i) has been actually solved already by Kraan and van Baal~\cite{KvB}.
Problem ii) is reduced to iii) in the self-dual background field~\cite{BC}
since for such fields $\Det(W_{\mu\nu})= \Det(-D^2)^4$, where $W_{\mu\nu}$
is the quadratic form for spin-1, isospin-1 quantum fluctuations and $D^2$ is
the covariant Laplace operator for spin-0, isospin-1 ghost fields. Symbolically,
one can write
\beq
{\rm KvBLL\;measure}=\int\!d({\rm collective\; coordinates})\cdot{\rm  Jacobian}\cdot
\Det^{-\frac{1}{2}}(W_{\mu\nu})\cdot\Det(-D^2)
\la{symbolic}\eeq
where the product of the last two factors is simply $\Det^{-1}(-D^2)$ in the
self-dual background. As usually, the functional determinants are  normalized
to free ones (with zero background fields) and UV regularized (we use the
standard Pauli--Villars method). Thus, to find the quantum weight of the
KvBLL caloron only the ghost determinant has to be computed. \\

To that end, we follow Zarembo~\cite{Zar} and find the derivative of
this determinant with respect to the holonomy or, more precisely, to
$\v\equiv\left.\sqrt{A_4^aA_4^a}\right|_{|\vec x|\to\infty}$. The derivative is
expressed through the Green function of the ghost field in the caloron background.
If a self-dual field is written in terms of the Atiah--Drinfeld--Hitchin--Manin
construction, and in the KvBLL case it basically is~\cite{KvB,LL}, the Green function
is generally known~\cite{Adler}-\cite{Nahm80} and we build it explicitly for
the KvBLL case. Therefore, we are able to find the derivative $\d\Det(-D^2)/\d\v$.
Next, we reconstruct the full determinant by integrating over $\v$
using the determinant for the trivial holonomy~\cite{GPY} as a boundary condition.
This determinant at $\v=0$ is still a non-trivial function of $\D $ and
the fact that we match it from the $\v\neq 0$ side is a serious check.
Actually we need only one overall constant factor from ref.~\cite{GPY} in order
to restore the full determinant at $\v\neq 0$, and we make a minor improvement
of the Gross--Pisarski--Yaffe calculation as we have computed the needed
constant analytically.  \\

Although all the above steps can be performed explicitly, at some point 
the equations become extremely lengthy -- typical expressions are several Mbytes
long and so far we have not managed to simplify them such that they would
fit into a paper. However, we are able to obtain compact analytical expressions
in the physically interesting case of large separation between dyons,
$\D\gg 1/T$. We have also used the exact formulae to check numerically some of
the intermediate formulae, in particular at $\v\to 0$.

If the separation is large in the temperature scale, $\D\gg 1/T$, the final result
for the quantum measure of the KvBLL caloron can be written down in terms of the
$3d$ positions of the two constituent $L,M$ dyons 
$\vec z_{1,2}$, their separation $\D=|\vec z_1-\vec z_2|$, the
asymptotic of $A_4$ at spatial infinity denoted by $\v\in[0,2\pi T]$
and $\bv= 2\pi T-\v\in[0,2\pi T]$, see \eq{Zfull}. We give here
a simpler expression obtained in the limit when the separation between dyons
is much larger than their core sizes:
\beqa
\nn
{\cal Z}_{\rm KvBLL}&= &\int d^3z_1\, d^3z_2\, T^6\, (2\pi)^{\frac{8}{3}}\,
C\, \left( \frac{8 \pi^2}{g^2} \right)^4
\left(\frac{\Lambda  e^{\gamma_E}}{4 \pi T}\right)^{\frac{22}{3}}
\left(\frac{\v}{2\pi T}\right)^{\frac{4 \v}{3\pi T}}
\left(\frac{\bv}{2\pi T}\right)^{\frac{4 \bv}{3\pi T}}\\
& \times &  \exp\left[-2\pi\,\D\,P''(\v)\right]\;\exp\left[-V^{(3)}P(\v)\right],
\eeqa
where the overall factor $C$ is a combination of universal constants; numerically
$C=1.031419972084$. $\Lambda$ is the scale parameter in the Pauli--Villars scheme;
the factor $g^{-8}$ is not renormalized at the one-loop level. \\

Since the caloron field has a constant $A_4$ component at spatial infinity,
it is suppressed by the same perturbative potential $P(\v)$ as given by \eq{Tpot}.
Its second derivative with respect to $\v$ is
$P''(\v)=\frac{1}{\pi^2 T}\left[v\!-\!\pi T\left(1\!-\!\frac{1}{\sqrt{3}}\right)\right]
\left[v\!-\!\pi T\left(1\!+\!\frac{1}{\sqrt{3}}\right)\right]$. If $\v$ is in the
ranges between 0 and $\pi T\left(1\!-\!\frac{1}{\sqrt{3}}\right)$ or between
$\pi T\left(1\!+\!\frac{1}{\sqrt{3}}\right)$ and $2\pi T$ (corresponding to the
holonomy not too far from the trivial, $0.787597\!<\!\half|\Tr L|\!<\!1$) 
the second derivative $P''(\v)$ is positive, and the $L$ and $M$ dyons experience 
a linear attractive potential. Integration over the separation $\D$ of dyons 
inside a caloron converges. We perform this integration in section VII, 
estimate the free energy of the caloron gas and conclude that trivial holonomy 
($\v=0,2\pi T$) may be unstable, despite the perturbative potential
energy $P(\v)$. In the complementary range  
$\pi T\left(1\!-\!\frac{1}{\sqrt{3}}\right)\!<\!\v\!<\!\pi T\left(1\!+\!\frac{1}{\sqrt{3}}\right)$ 
(or $\half|\Tr L|\!<\!0.787597$), $P''(\v)$ is negative (see Fig.~2), and dyons experience
a strong linear-rising repulsion. It means that for these values of $\v$,
integration over the dyon separations diverges: calorons with holonomy
far from trivial ``ionize'' into separate dyons.

\section{The KvBLL caloron solution}

\begin{figure}[t]
\centerline{
\epsfxsize=0.4\textwidth
\epsfbox{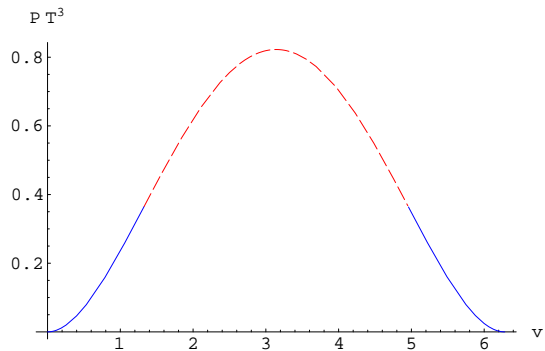}}
\caption{Potential energy as function of $\v/T$.
Two minima correspond to $\half\Tr L=\pm 1$, the maximum corresponds to $\Tr L=0$.
The range of the holonomy where dyons experience repulsion is shown in dashing.
\label{fig:pot}}
\end{figure}

Although the construction of the self-dual solution with
non-trivial holonomy has been fully performed independently by
Kraan and van Baal~\cite{KvB} and Lee and Lu~\cite{LL} we have
found it more convenient for our purposes to use the gauge
convention and the formalism of Kraan and van Baal (KvB) whose
notations we follow in this paper.

The key quantities characterizing the KvBLL solution for a general
$SU(N)$ gauge group are the $N\!-\!1$ gauge-invariant eigenvalues
of the Polyakov line \ur{Pol1} at spatial infinity. For the
$SU(2)$ gauge group to which we restrict ourselves in this paper,
it is just one quantity, e.g. $\Tr\,L$, \eq{TrL}. In a gauge where
$A_4$ is static and diagonal at spatial infinity, i.e.
$\left.A_4\right|_{{\vec x}\to\infty}= i\v\frac{\tau_3}{2}$, it is
this asymptotic value $\v$ which characterizes the caloron
solution in the first place. We shall also use the complementary
quantity $\bv \equiv 2\pi T-\v$. Their relation to parameters
$\omega, \bar\omega$ introduced by KvB~\cite{KvB} is
$\omega= \frac{\v}{4\pi T},\; \bar\omega= \frac{\bv}{4\pi  T}= \half-\omega$.
Both $\v$ and $\bv$ vary from 0 to $2\pi T$. At $\v= 0,2\pi T$ the  holonomy is said to
be `trivial', and the KvBLL caloron reduces to that of Harrington and  Shepard~\cite{HS}.

There are, of course, many ways to parametrize the caloron
solution. Keeping in mind that we shall be mostly interested in
the case of widely separated dyon constituents, we shall
parametrize the solution in terms of the coordinates of the dyons'
`centers' (we call constituent dyons L and M according to the
classification in ref.~\cite{DPSUSY}):
\beqa
\nn
L\;{\rm dyon}:\quad &&\vec z_1= -\frac{2\omega{\overrightarrow  \D}}{T},\\
\nn
M\;{\rm dyon}:\quad &&\vec z_2= \frac{2\bar\omega{\overrightarrow  \D}}{T},\\
\nn {\rm dyon\; separation}: &&\vec z_2-\vec z_1= {\overrightarrow  \D},\quad
\quad |\D |= \pi T\,\rho^2,
\eeqa
where
$\rho$ is the parameter used by KvB; it becomes the size of the
instanton at $\v\to 0$. We introduce the distances from the
`observation point' $\vec x$ to the dyon centers,
\beqa
\nn
&&\vec r = \vec x-\vec z_1=\vec x + 2\omega\overrightarrow{\D},\qquad r= |\vec r|,\\
&&\vec s = \vec x-\vec z_2=\vec x - 2\bar\omega\overrightarrow{\D},\qquad s= |\vec s|.
\eeqa
Henceforth we measure all dimensional quantities in units of temperature
for brevity and restore $T$ explicitly only in the final results.

The KvBLL caloron field in the fundamental representation
is~\cite{KvB} (we choose the separation between dyons to be in the
third spatial direction, ${\overrightarrow \D}= \D\vec e_3$):
\beq
A_\mu= \delta_{\mu,4}\,i\v\frac{\tau_3}{2}
+\frac{i}{2}\bar\eta^3_{\mu\nu}\tau_3\partial_\nu\ln\Phi
+\frac{i}{2}\Mphi\;{\rm  Re}\left[(\bar\eta^1_{\mu\nu}-i\bar\eta^2_{\mu\nu})
(\tau_1+i\tau_2)(\partial_\nu+i\v\delta_{\nu,4})\tilde\chi\right] \,,
\label{APvB}\eeq
where $\tau_i$ are Pauli matrices, $\bar\eta^a_{\mu\nu}$ are 't~Hooft's
symbols~\cite{tHooft} with $\bar\eta^a_{ij}= \epsilon_{aij}$ and
$\bar\eta^a_{4\nu}= -\bar\eta^a_{\nu 4}= i\delta_{a\nu}$. ``Re''
means $2{\rm Re}(W)\equiv W+W^\dag$ and the functions used are
\beqa
\nn
&&\hat\psi= -\cos(2\pi x_4)+\bchd\chd+\frac{\vec r\vec  s}{2rs}\bshd\shd,\qquad
\psi= \hat\psi+\frac{\D^2}{rs}\bshd\shd+\frac{\D}{s}\shd\bchd+\frac{\D}{
r}\bshd\chd,\\
&&\tilde\chi= \frac{\D}{\psi}\left(e^{-2\pi  ix_4}\frac{\shd}{s}+\frac{\bshd}{r}\right),
\qquad\Mphi= \frac{\psi}{\hat\psi}\,.
\label{eq:psihat}\eeqa
We have introduced short-hand notations for hyperbolic functions:
\beq
\shd\equiv\sinh(s\v),\qquad \chd\equiv\cosh(s\v),\qquad
\bshd\equiv\sinh(r\bv),\qquad \bchd\equiv\cosh(r\bv) \,.
\eeq
The first term in \ur{APvB} corresponds to
a constant $A_4$ component at spatial infinity ($ A_4 \approx  i\v\frac{\tau_3}{2} $)
and gives rise to the non-trivial holonomy.
One can see that $A_\mu$ is periodic in time with period $1$
(since we have chosen the temperature to be equal to unity). A
useful formula for the field strength squared is~\cite{KvB}
\beq
\Tr\,F_{\mu\nu}F_{\mu\nu}=\partial^2\partial^2\log\psi.
\la{FF}\eeq

In the situation when the separation between dyons $\D$ is large
compared to both their core sizes $\frac{1}{\v}$ (M) and
$\frac{1}{\bv}$ (L), the caloron field can be approximated by the
sum of individual BPS dyons, see Figs.~1,3 (left) and Fig.~4. 
We give below the field inside the cores and far away from both cores.

\begin{figure}[t]
\centerline{
\epsfxsize=0.6\textwidth
\epsfbox{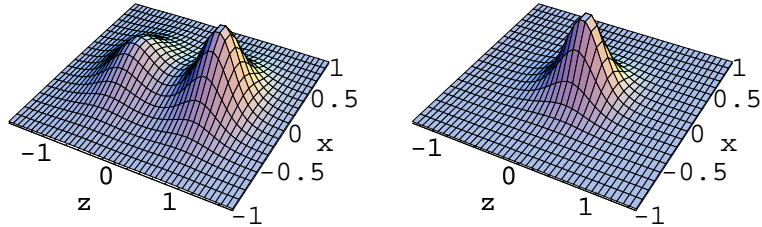}}
\caption{\label{fig:adp2} The action density of the KvBLL caloron as
function of $z,x$ at fixed $t=y=0$. At large separations $r_{12}$ the caloron
is a superposition of two BPS dyon solutions (left, $r_{12}=1.5/T$). At
small separations they merge (right, $r_{12}=0.6/T$). The caloron
parameters are the same as in Fig.~1.}
\end{figure}

\subsection{Inside dyon cores}

In the vicinity of the L dyon center $\vec z_1$ and far away from
the M dyon ($s \v \gg 1$) the field becomes that of the L dyon.
It is instructive to write it in spherical coordinates centered at $\vec
z_1$.
In the `stringy' gauge~\cite{DPSUSY} in which the $A_4$ component is
constant and diagonal at spatial infinity, the L dyon field is
\beqa
\nn A_4^L&=&\frac{i\tau_3}{2}\left(\frac{1}{r}+2\pi-\bv\coth(\bv r)\right),\;\;\;A^L_r=0,\\
\nn A^L_\theta &=& i\bv\frac{-\sin(2\pi x_4-\phi)\;\tau_1
+\cos(2\pi x_4-\phi)\tau_2}{2\sinh(\bv r)},\\
A^L_\phi &=& i\bv\frac{\cos(2\pi x_4-\phi)\;\tau_1+\sin(2\pi
x_4-\phi)\tau_2}{2\sinh(\bv r)}-i\tau_3\frac{\tan(\theta/2)}{2r}\;.
\la{Ldyon}\eeqa
Here $A_\theta$, for example, is the projection of $\vec A$ onto the direction
$\vec{n}_\theta=(\cos\theta\cos\phi,\cos\theta\sin\phi,-\sin\theta)$.
The $\phi$ component has a string singularity along the $z$ axis
going in the positive direction. Notice that inside the core region ($\bv r\leq 1$)
the field is time-dependent, although the action density is static.
At large distances from the L dyon center, i.e. far outside the core
one neglects exponentially small terms ${\cal O}\left(e^{-\bv r}\right)$
and the surviving components are
\beqa
\nn
A^L_4 &\stackrel{r\to\infty}{\longrightarrow}&\left(\v+\frac{1}{r}\right)\,\frac{i\tau^3}{2},\\
\nn
A^L_\phi &\stackrel{r\to\infty}{\longrightarrow}& 
-\frac{{\rm tan}\frac{\theta}{2}}{r}\,\frac{i\tau^3}{2}
\eeqa
corresponding to the radial electric and magnetic field components
\beq
E^L_r= B^L_r\stackrel{r\to\infty}{\longrightarrow}
-\frac{1}{r^2}\,\frac{ i\tau^3}{2}.
\la{Ldyon_as}\eeq
This Coulomb-type behavior of both the electric and magnetic fields
prompts the name `dyon'.

Similarly, in the vicinity of the M dyon and far away from the L
dyon ($r \bv \gg1$) the field becomes that of the M dyon, which we
write in spherical coordinates centered at $\vec z_2$:
\[
A_4^M =\frac{i\tau_3}{2}\left(\v\coth(\v s)-\frac{1}{s}\right),\quad
A_\theta^M=\v\frac{\sin\!\phi\;\tau_1-\cos\!\phi\;\tau_2}{2i\sinh(\v s)}\,,
\]
\beq
A_r^M =0,\quad A_\phi^M=\v\frac{\cos\!\phi\;\tau_1+\sin\!\phi\;\tau_2}{2i\sinh(\v s)}
+i\tau_3\frac{\tan(\theta/2)}{2s},
\la{Mdyon}\eeq
whose asymptotics is
\beqa
\nn
A^M_4 &\stackrel{r\to\infty}{\longrightarrow}& \left(\v-\frac{1}{s}\right)\,\frac{i\tau^3}{2},\\
\nn A^M_\phi &\stackrel{r\to\infty}{\longrightarrow}& \frac{{\rm  tan}\frac{\theta}{2}}{s}\,
\frac{i\tau^3}{2},\\
E^M_r=B^M_r &\stackrel{r\to\infty}{\longrightarrow}&
\frac{1}{s^2}\,\frac{i\tau^3}{2}.
\label{Mdyon_as}\eeqa
We see that in both cases the L,M fields become Abelian at large distances, 
corresponding to (electric, magnetic) charges $(-,-)$ and $(+,+)$, respectively.
The corrections to the fields \urs{Ldyon}{Mdyon} are hence of the
order of $1/\D$ arising from the presence of the other dyon.

\subsection{Far away from dyon cores}

Far away from both dyon cores ($r \bv\gg 1,\;s\v \gg 1$; note that
it does not necessarily imply large separations -- the dyons may
even be overlapping) one can neglect both types of exponentially
small terms, ${\cal O}\left(e^{-r\bv }\right)$ and ${\cal  O}\left(e^{-s\v}\right)$.
With exponential accuracy the function $\tilde\chi$ in \eq{eq:psihat} is zero,
and the KvBLL field \ur{APvB} becomes Abelian~\cite{KvB}:
\beq
A_\mu^{\rm as}= i\frac{\tau_3}{2}\left(\delta_{\mu 4}\v
+\bar\eta^3_{\mu\nu}\partial_\nu\ln\,\Phi_{\rm as}\right),
\la{Aas}\eeq
where $\Phi_{\rm as}$ is the function $\Phi$ of \eq{eq:psihat} evaluated
with the exponential precision:
\beq
\Phi_{\rm as}= \frac{r+s+\D}{r+s-\D}= \frac{s-s_3}{r-r_3}\;\;
{\rm if}\;\;{\overrightarrow \D}= \D\vec e_3.
\la{Phias}\eeq
It is interesting that, despite being Abelian, the asymptotic field \ur{Aas} retains its
self-duality. This is because the $3^{\rm d}$ color component of the electric field is
\beq
\nn
E_i^3=\partial_iA_4^3=\partial_i\partial_3\ln\,\Phi_{\rm as}
\eeq
while the magnetic field is
\beq
\nn
B_i^3=\epsilon_{ijk}\partial_jA_k^3= \partial_i\partial_3\ln\,\Phi_{\rm as}
-\delta_{i3}\partial^2\ln\,\Phi_{\rm as},
\eeq
where the last term is zero, except on the line connecting the dyon centers where
it is singular; however, this singularity is an artifact of the exponential
approximation used. Explicit evaluation of \eq{Aas} gives the following nonzero
components of the $A_\mu$ field far away from both dyon centers:
\beqa
\la{A4as}
&&A_4^{\rm  as}= \frac{i\tau_3}{2}\left(\v+\frac{1}{r}-\frac{1}{s}\right),\\
\la{Aphias}
&&A_\varphi^{\rm  as}= -\frac{i\tau_3}{2}\left(\frac{1}{r}+\frac{1}{s}\right)
\sqrt{\frac{(\D-r+s)(\D+r-s)}{(\D+r+s)(r+s-\D)}}\;.
\eeqa
In particular, far away from both dyons, $A_4$ is the Coulomb field
of two opposite charges.

\section{The scheme for computing Det${\bf (-D^2)}$}

As explained in section I, to find the quantum weight of the KvBLL
caloron, one needs to calculate the small oscillation determinant,
${\rm Det}(-D^2)$, where $D_\mu= \d_\mu+A_\mu$ and $A_\mu$ is the
caloron field \ur{APvB}. Instead of computing the determinant
directly, we first evaluate its derivative with respect to the
holonomy $\v$, and then integrate the derivative using the known
determinant at $\v= 0$~\cite{GPY} as a boundary condition.

If the background field $A_\mu$ depends on some parameter ${\cal P}$,
a general formula for the derivative of the determinant with respect
to such parameter is
\beq
\frac{\partial\,\log {\rm Det}(-D^2[A])}{\partial {\cal P}}
= \!-\!\int d^4x\,\Tr\left(\partial_{\cal P} A_\mu\, J_\mu\right),
\label{dvDet}\eeq
where $J_\mu$ is the vacuum current in the external background,  determined
by the Green function:
\beq
J^{ab}_\mu\!\equiv\!\left.(\delta^a_c\delta^b_d\d_x\!
-\!\delta^a_c\delta^b_d\d_y\!+\!A^{ac}\delta^b_d\!
+\!A^{db}\delta^a_c) {\cal G}^{cd}(x,y)\right|_{y= x}\qquad
{\rm or\;simply}\quad J_\mu\equiv \overrightarrow{D}_\mu {\cal G}
+{\cal G} \overleftarrow{D}_\mu.
\label{defJ}\eeq
Here $\cal G$ is the Green function or the propagator of spin-0,  isospin-1
particle in the given background $A_\mu$ defined by
\beq
-D^2_xG(x,y)= \delta^{(4)}(x-y)
\la{Gdef}\eeq
and, in the case of nonzero temperatures, being periodic in time,  meaning that
\beq
{\cal G}(x,y)= \sum_{n= -\infty}^{\infty}G(x_4,{\vec x};y_4+n,{\vec  y}).
\la{greenP}\eeq
\Eq{dvDet} can be easily verified by differentiating the identity
$\log{\rm Det}(-D^2)= \Tr\log(-D^2)$ \footnote{Generally speaking, there is
a surface term in \eq{dvDet} arising from integrating by  parts~\cite{Zar}, which we
ignore here since the caloron field, contrary to the single dyon's
one considered by Zarembo~\cite{Zar}, decays fast enough at
spatial infinity. We take this opportunity to say that we have
learned much from Zarembo's paper. However, his consideration of
dyons with high charge $k$ only didn't allow him to observe the
subleading in $k$ infrared divergence of the single dyon
determinant, which is the essence of the problem with individual
dyons, as contrasted to charge-neutral calorons. In addition,
because dyons have to be always regularized by putting them in a
finite-size box, the theorem on the relation between spin-1 and
spin-0 determinants (section I) is generally violated. This is one
of the reasons we consider well-behaved calorons rather than
individual dyons, although they are more ``elementary".}. The
background field $A_\mu$ in \eq{dvDet} is taken in the adjoint
representation, as is the trace. Hence, if the periodic propagator
${\cal G}$ is known, \eq{dvDet} becomes a powerful tool for
computing quantum determinants. Specifically, we take ${\cal P}= \v$
as the parameter for differentiating the determinant, and
there is no problem in finding $\partial_{\v}A_\mu$ for the caloron 
field \ur{APvB}.

The Green functions in self-dual backgrounds are generally
known~\cite{CWS,Nahm80} and are built in terms of the
Atiah--Drinfeld--Hitchin--Manin (ADHM) construction~\cite{ADHM}
for the given self-dual field. A subtlety appearing at nonzero
temperatures is that the Green function is defined by \eq{Gdef} in
the Euclidean $\re^4$ space where the topological charge is
infinite because of the infinite number of repeated strips in the
compactified time direction, whereas one actually needs an
explicitly periodic propagator \ur{greenP}. To overcome this
nuisance, Nahm~\cite{Nahm80} suggested to pass on to the Fourier
transforms of the infinite-range subscripts in the ADHM
construction. We perform this program explicitly in Appendix A,
first for the single dyon field and then for the KvBLL caloron. In
this way, we get the finite-dimensional ADHM construction both for
the dyon and the caloron, with very simple periodicity properties.
Using it, we construct explicitly periodic propagators in Appendix
B, also first for the dyon and then for the caloron case.
For the KvBLL caloron it was not known previously.
Using the obtained periodic propagators, in Appendix C we calculate
the exact vacuum current \ur{defJ} for the dyon, and in Appendix D
we evaluate the vacuum current in the caloron background, with the help of the
regularization carried out in Appendix E.

\begin{figure}[t]
\centerline{
\epsfxsize=0.5\textwidth
\epsfbox{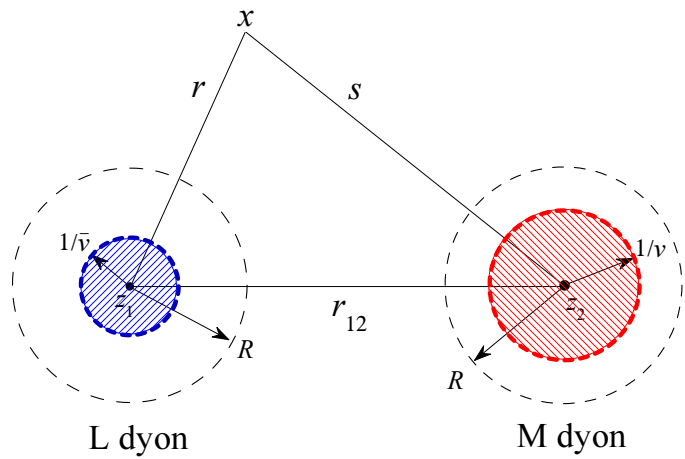}}
\caption{\label{fig:LMdyons} Three regions of integration for well separated dyons.}
\end{figure}

Although there is no principle difficulty in doing all
calculations exactly for the whole caloron moduli space, at some
point we loose the capacity of performing analytical calculations
for the simple reason that expressions become too long, and so far
we have not been able to put them into a manageable form in a
general case. Therefore, we have to adopt a more subtle attitude.
First of all we restrict ourselves to the part of the moduli space
corresponding to large separations between dyons ($\D\gg 1$).
Physically, it seems to be the most interesting case, see 
section~I. Furthermore, at the first stage we take $\D\v,\D\bv\gg 1$,
meaning that the dyons are well separated and do not
overlap since the separation is then much bigger than the core
sizes, see Figs.~1,3 (left). In this case, the vacuum current
$J_\mu$ \ur{defJ} becomes that of single dyons inside the spheres
of some radius R surrounding the dyon centers, such that
$\frac{1}{\v},\frac{1}{\bv}\ll R \ll \D$, and outside these
spheres it can be computed analytically with exponential
precision, in correspondence with subsection II.B, see Fig.~4. Adding up the
contributions of the regions near two dyons and of the far-away
region, we get $d\Det(-D^2)/d\v$ for well-separated dyons.
Integrating it over $\v$ we obtain the determinant itself up to a
constant and possible $1/\D$ terms.

This is already an interesting result by itself, however, we would
like to compute the constant, which can be done by matching our
calculation with that for the trivial caloron at $\v=0$. It means
that we have to extend the domain of applicability to $\D\v=\O(1)$
(or $\D\bv=\O(1)$) implying overlapping dyons, presented in Figs.~1,3
(right). To make this extension, we `guess' the analytical
expression which would interpolate between $\D\v\gg 1$ where
the determinant is already computed and $\D\v\ll 1$ where matching
with the Gross--Pisarski--Yaffe (GPY) calculation~\cite{GPY} can be performed.
At this point it becomes very helpful that we possess the exact
vacuum current for the caloron, which, although too long
to be put on paper, is nevertheless affordable for numerical
evaluation (and can be provided on request). We check our analytical
`guess' to the accuracy better than one millionth. In this way
we obtain the determinant up to an overall constant factor for any
$\v,\bv$ with the only restriction that $\D\gg 1$. This constant
factor is then read off from the GPY calculation~\cite{GPY}.

Finally, we compute the $1/\D$ and $\log\D/\D$ corrections in the
$\Det(-D^2)$, which turn out to be quite non-trivial.

\section{Det${\bf (-D^2)}$ for well separated dyons}

The L,M dyon cores have the sizes $\frac{1}{\bv}$ and $\frac{1}{\v}$, respectively, and in
this section we consider the case of well-separated dyons, meaning that the distance between
the two centers is much greater than the core sizes, $\D\gg \frac{1}{\v},\;\frac{1}{\bv}$.
This situation is depicted in Figs.~1,3 (left). The two dyons are static in time, so that
$\d\,\log\Det(-D^2)/\d\v$ \ur{dvDet} becomes an integral over $3d$  space, times $1/T$ set to unity.
We divide the $3d$ volume into to three regions (Fig.~4): i) a ball of radius R centered at the center
of the M dyon, ii) a ball of radius R centered at the L dyon, iii) the rest of the space,
with two balls removed. The radius R is chosen such that it is much larger than the dyon cores
but much less than the separation: $\D\gg R\gg \frac{1}{\v},\;\frac{1}{\bv}$. Summing up
the contributions from the three regions of space, we are satisfied to observe
that the result does not depend on the intermediate radius $R$.

\subsection{Det${\bf (-D^2)}$ for a single dyon}

In region i) the KvBLL caloron field can be approximated by the M dyon field \ur{Mdyon},
and the vacuum current by that inside a single dyon, both with the $\O(1/\D)$ accuracy.
We make a more precise calculation, including the $\O(1/\D)$ terms, in section V.
The single-dyon vacuum current is calculated in Appendix C. Adding up the three
parts of the vacuum current denoted there as $J_\mu^{\rm s,r,m}$ we obtain the full
isospin-1 vacuum current (in the `stringy' gauge)
\beqa
\la{monopoleCurrent}
\nonumber J_r&= &0,\\
\nonumber J_\phi&= &-\frac{i\v(\sin(\phi)T_2+\cos(\phi)T_1)(1-s\v\coth(s\v))^2}{24\pi^2 s^2\sinh(s\v)},\\
\nonumber J_\theta&= &-\frac{i\v(\sin(\phi)T_1-\cos(\phi)T_2)(1-s\v\coth(s\v))^2}{24\pi^2 s^2\sinh(s\v)},\\
\nonumber J_4&= &-iT_3\left[\frac{(1-s\v \coth(s\v))^3}{6\pi^2s^3}+\frac{1-s\v\coth(s\v)}{3s}
+\frac{\coth(s\v )(1-s\v \coth(s\v ))^2}{2\pi s^2}\right]\;,
\eeqa
where ${(T_c)}^{ab}\equiv i\varepsilon^{acb}$ are the isospin-1 generators. We contract
$J_\mu$ \ur{monopoleCurrent} with $dA_\mu/d\v$ from \eq{Mdyon} according  to \eq{dvDet}.
After taking the matrix trace, the integrand in \eq{dvDet} becomes spherically
symmetric:
\beqa
\nn
\Tr\left[\d_\v A_\mu J_\mu\right]&=&\frac{2}{3s}
\left(\coth(s\v)-s\v +\frac{s\v \left(s\v\coth(s\v)-2\right)}{\sinh^2(s\v)}\right) \\
&-&\frac{{\left(s\v\coth(s\v)-1\right)}^3\left(\sinh(2 s\v)-3s\v\right)}{6\pi^2\sinh^2(s\v)s^3}
+\frac{{\left(s\v\coth(s\v)-1\right)}^2\left(\sinh(2 s\v)-2s\v\right)}{2\pi\sinh^2(s\v)\tanh(s\v)s^2};
\la{integrand1}\eeqa
It has to be integrated over the spherical box of radius R. Fortunately, we are able
to perform the integration analytically. The result for the M dyon is
\beqa
\nn
\frac{\d\,\log\,\Det(-D^2[{\rm M\,dyon}])}{\d \v}
&= &-\int_0^R\! ds\,\Tr\left[\d_\v A_\mu J_\mu\right] 4\pi s^2
= -\frac{24(\gamma_E-\log\pi)+53+\frac{2}{3}\pi^2}{18\pi}+\frac{1}{\v}\\
&+&\frac{4\pi R^3}{3}P'(\v)-2 \pi R^2\,P''(\v)+2\pi R\,P'''(\v)-\frac{4}{3\pi}\log(R\v).
\label{intM}\eeqa
As we see, it is badly infrared divergent, as it depends on the box radius $R$.
Here $P(q)$ is the potential energy (see \eq{Tpot})
\beqa
\nn
&&P(q)=\left[\frac{\pi^2}{12}\left(\frac{q}{\pi}-2\right)^2\left(\frac{q}{\pi}\right)^2\right],\qquad
P'(q)=\frac{1}{3\pi^2}q(\pi-q)(2\pi-q),\\
\nn
&&P''(q)=\frac{1}{3\pi^2}(3q^2-6\pi q+2\pi^2),\qquad
P'''(q)=\frac{2}{\pi^2}(q-\pi),\qquad P^{\rm IV}(q)=\frac{2}{\pi^2}.
\la{pot}\eeqa
The IR-divergent terms arise from the asymptotics of the integrand.
Neglecting exponentially small terms $e^{-s\v}$ in \eq{integrand1} we have
\beqa
\nn
-\Tr\left[\d_\v A_\mu J_\mu\right]&\simeq & -4\left[\frac{(1-s\v)^3}{12\pi^2s^3}
+\frac{(1-s\v)^2}{4\pi s^2}+\frac{1-s\v}{6 s}\right]\\
\nn
&&= P'\left(\v-\frac{1}{s}\right)=P'(\v)-P''(\v)\frac{1}{s}+\frac{1}{2}P'''(\v)\frac{1}{s^2}
-\frac{1}{6}P^{\rm IV}(\v)\frac{1}{s^3}.
\label{smep}\eeqa
Integrating it over the sphere of radius R one gets the IR-divergent terms (the second
line in \eq{intM}).

The fact that the IR-divergent part of $d\Det(-D^2)/d\v$ is directly related to the potential
energy $P(A_4)$ is not accidental. At large distances the field of the dyon becomes a
slowly varying Coulomb field, see \eq{Mdyon_as}. Therefore, the determinant can be generically
expanded in the covariant derivatives of the background field~\cite{DO,RA} with the potential
energy $P(A_4)$ being its leading zero-derivative term. The nontrivial  fact, however, is that
with exponential precision the vacuum current is related to the variation of solely the
leading term in the covariant derivative expansion of the effective action with no contribution from
any of the subleading terms. This is a specific property of self-dual fields, and we observe it
also in the following subsection.

\subsection{Contribution from the far-away region}

We now compute the contribution to $\d\Det(-D^2)/\d\v$ from the region of space far away
from both dyon centers. With exponential accuracy (meaning neglecting terms of the order of
$e^{-r\bv}$ and $e^{-s\v}$) the KvBLL caloron field is given by \eqs{A4as}{Aphias}, and
only the $A_4$ component depends (trivially) on $\v$.
The caloron vacuum current with the same exponential accuracy is calculated in Appendix D.
Combining the results given by \eqs{j4cals}{jphicals} and \eqs{J4calr}{Jphicalr}
we see that $J_\varphi= 0$ and for $J_4$ we have
\begin{widetext}
\beqa
\nn
J_4&= &\frac{i T_3}{2}\left\{
\left[\frac{1}{3\pi^2} \left(\frac{1}{r}-\frac{1}{s}\right)^3
-\frac{1}{\pi} \left(\frac{1}{r}-\frac{1}{s}\right)^2
+\frac{2}{3} \left(\frac{1}{r}-\frac{1}{s}\right) \right]
+\left[\frac{4}{\pi} \left(\frac{1}{r}-\frac{1}{s}\right)^2
-8 \left(\frac{1}{r}-\frac{1}{s}\right) +  \frac{8\pi}{3}\right]\omega\right. \\
&+&\left.\left[16 \left(\frac{1}{r}-\frac{1}{s}\right) - 16\pi  \right]\omega^2
+\frac{64\pi}{3}\,\omega^3 \right\}\,.
\label{farCurrent}\eeqa
\end{widetext}
We remind the reader that $r,s$ are distances from M,L dyon centers $\vec z_{1,2}$
and that $\omega=\v/(4\pi)$.
It is interesting that the separation $\D= |z_1-z_2|$ does not appear
explicitly in the current. Moreover, it can be again written through the
potential energy $P(A_4)$:
\beq
J_4=\frac{1}{2} iT_3\left.P'(q)\right|_{q=\v+1/r-1/s}\,.
\la{J4asP}\eeq
Therefore, in the far-away region one obtains
\beq
-\Tr\left[\d_\v A_\mu J_\mu\right]= P'\left(\v+\frac{1}{r}-\frac{1}{s}\right) \;.
\la{integrand_far}\eeq

We have now to integrate \eq{integrand_far} over the whole $3d$ space
with two spheres of radius $R$ surrounding the dyon centers removed:
\beqa
\la{far1}
-\int\!d^4x\,\Tr\left[\d_\v A_\mu J_\mu\right]
&=&\int\!d^3x\,P'\left(\v+\frac{1}{r}-\frac{1}{s}\right)\\
\nn
&= &P'(\v)\!\int\!d^3x+P''(\v)\!\int\!\!d^3x\left(\frac{1}{r}-\frac{1}{s}\right)
+\frac{1}{2}P'''(\v)\!\int\!\!d^3x\left(\frac{1}{r}-\frac{1}{s}\right)^2
+\frac{1}{6}P^{\rm  IV}(\v)\!\int\!\!d^3x\left(\frac{1}{r}-\frac{1}{s}\right)^3.
\eeqa
The first integral in \eq{far1} is the $3d$ volume $V$, minus the volume of two spheres,
$V-2\frac{4\pi}{3}R^3$. The second integral is zero by symmetry between the two
centers, and so is the last one. The only non-trivial integral is
\beq
\int\!d^3x \,\left(\frac{1}{r}-\frac{1}{s}\right)^2
=4\pi \D-8\pi R+{\cal O}\left(\frac{R^2}{\D}\right).
\la{Norditaboard}\eeq
Therefore, the contribution from the region far from both dyon centers is
\beq
\left.\frac{\d\,\log\,\Det(-D^2)}{\d\v}\right|_{\rm far}
= P'(\v)\left(V-2\frac{4\pi}{3}R^3\right)+\frac{1}{2}P'''(\v)\left(4\pi\D-8\pi R\right).
\la{far2}\eeq

\subsection{Combining all three regions}

We now add up the contributions to $\d\log\,\Det(-D^2)/\d\v$ from
the regions surrounding the two dyons and from the far-away region.
Since the contribution of the L  dyon is the same as that of the M dyon
with the replacement $\v\to\bv$ and since  $\d/\d\v=-\d/\d\bv$, when
adding up contributions of L,M core regions we have to  antisymmetrize
in $\v\leftrightarrow\bv$. It should be noted that $P(\v)$ and $P''(\v)$
are symmetric under this interchange, while $P'(\v)$ and $P'''(\v)$ are
antisymmetric. Therefore, the combined contribution of both cores is, from \eq{intM},
\beq
\la{cores}
\left.\frac{\d\,\log\,\Det(-D^2)}{\d\v}\right|_{\rm cores}
= 2 P'(\v)\frac{4\pi}{3}R^3+\frac{1}{2}P'''(\v)\,8\pi R
+\frac{1}{\v}-\frac{1}{\bv}-\frac{4}{3\pi}\ln\left(\frac{\v}{\bv}\right).
\eeq
Adding it up with the contribution from the far-away region, \eq{far2}, we obtain the final
result which is independent of the intermediate radius $R$ used to separate the regions:
\beq
\frac{\d\,\log\,\Det(-D^2)}{\d\v} = P'(\v)\,V
+P'''(\v)\,2\pi \D +\frac{1}{\v}-\frac{1}{\bv}-\frac{4}{3\pi}\ln\left(\frac{\v}{\bv}\right).
\la{largesep1}\eeq
This equation can be easily integrated over $\v$ up to a constant which
in fact can be a function of the separation $\D$:
\beq
\log\,\Det(-D^2)=P(\v)\,V+P''(\v)\,2\pi \D
+\left(\!1-\frac{4 \v}{3\pi}\!\right)\log(\v)
+\left(\!1-\frac{4 \bv}{3\pi}\!\right)\log(\bv)+ f(\D) \;.
\la{largesep2}\eeq
Since in the above calculation of the determinant for well-separated dyons we
have neglected the Coulomb field of one dyon inside the core region of the other,
we expect that the unknown function $f(\D)=\O(1/\D)+c$, where
$c$ is the true integration constant. Our next aim will be to find it.
The $\O(1/\D)$ corrections will be found later.

\section{Matching with the determinant with trivial holonomy}

To find the integration constant, one needs to know the value of the determinant
at $\v\!=\!0$ (or $\bv\!=\!0$) where the KvBLL caloron with non-trivial holonomy reduces
to the Harrington--Shepard caloron with a trivial one and for which the
determinant has been computed by GPY~\cite{GPY}. Before we match our determinant
at $\v\!\neq \!0$ with that at $\v\!= \!0$ let us recall the GPY result.

\subsection{Det${\bf (-D^2)}$ at $\v=0$}

The $\v\!=\!0$ periodic instanton is traditionally parameterized by the instanton size
$\rho$. It is known~\cite{Rossi,GPY} that the periodic instanton can be viewed
as a mix of two BPS monopoles one of which has an infinite size. It becomes especially
clear in the KvBLL construction~\cite{KvB,LL} where the size of one of the dyons becomes
infinite as $\v\to 0$, see section II. Despite one dyon being infinitely large,
one can still continue to parametrize a caloron by the distance $\D$ between dyon centers,
with $\rho= \sqrt{\D/\pi}$. Since our determinant \ur{largesep2} is given in terms of $\D$
we have first of all to rewrite the GPY determinant in terms of $\D$, too. Actually,
GPY have interpolated the determinant in the whole range of $\rho$ (hence $\D$) but we
shall be interested only in the limit $\D\gg 1$. In this range the GPY result reads:
\beqa
\la{GPY1}
\left.\log \Det(-D^2)\right|_{\v=0,\;T=1}
&= &\left.\log \Det(-D^2)\right|_{\v=0,\;T=0}+\frac{4}{3}\pi \D
-\frac{4}{3}\log \D+c_0+\O\left(\frac{1}{\D}\right), \\
\nn
c_0&= &\frac{8}{9}-\frac{8\,\gamma_E}{3}-\frac{2\,\pi^2}{27}+\frac{4\,\log \pi}{3}.
\eeqa
We have made here a small improvement as compared to ref.~\cite{GPY}, namely
i) we have checked that the correction is of the order of $1/\D$ basing on an intermediate
exact formula, ii) we have also managed to get an exact analytical expression for the constant.

The zero-temperature determinant is that for the standard BPST instanton~\cite{tHooft,Bernard}:
\beqa
\la{Dinst}
\left.\log \Det(-D^2)\right|_{\v= 0,\;T= 0}&= & \frac{2}{3} \log \mu
+ \frac{1}{3} \log\left(\frac{\D}{\pi}\right) +\alpha(1),\\
\alpha(1)&= &\frac{2\gamma_E}{3}- \frac{16}{9} +\frac{\log 2}{3} +  \frac{2\log (2\pi )}{3}
-\frac{4\zeta'(2)}{{\pi }^2}\;,
\eeqa
where it is implied that the determinant is regularized by the Pauli--Villars method
and $\mu$ is the Pauli--Villars mass, see section VI.A.
Combining \eqs{GPY1}{Dinst} one obtains
\beq
\left.\log \Det[-D^2]\right|_{\v= 0,\;T= 1}= \frac{4}{3}\pi \D
- \log\D +\frac{2}{3}\log \mu+c_1+\O\left(\frac{1}{\D}\right),
\label{GPY2}\eeq
where
\beqa
c_1= \log 2+\frac{5}{3} \log \pi-\frac{8}{9}-2\,\gamma_E
-\frac{2\,\pi^2}{27}-\frac{4\,\zeta'(2)}{{\pi }^2}= 0.206602292859\,.
\la{c1}\eeqa
We notice that $P(\v)\to 0$ and $P''(\v)\to \frac{2}{3}$ at $\v\to 0$,
therefore the first two terms in \eq{largesep2} become exactly equal to
the first term in \eq{GPY2}. At the same time, the last two terms in \eq{largesep2}
become $\log\v-\frac{5}{3}\log(2\pi)+c$ which is formally singular at  $\v\to 0$
and does not match the $-\!\log \D$ in \eq{GPY2}. The reason is, \eq{largesep2}
has been derived assuming $\D\gg \frac{1}{\v},\frac{1}{\bv}$ and one cannot
take the limit $\v\to 0$ in that expression without taking simultaneously $\D\to\infty$.
In order to match the determinant at $\v= 0$ one needs to extend
\eq{largesep2} to arbitrary values of $\v\D$. As we shall see,
it will be important for the matching that $\log \D$ has the coefficient $-1$.

\subsection{Extending the result to arbitrary values of $\v\D$}

Let us take a fixed but large value of the dyon separation $\D \gg 1$ such that
both \eq{largesep2} and \eq{GPY2} are valid. Actually, our aim is to integrate
the exact expression for the derivative of the determinant
\beq
\label{aim}
\partial_\v \log \Det(-D^2)= \int \wp(x)d^4x ,\qquad
\wp(x)\equiv -\Tr\left[\partial_\v A_\mu J^\mu(x,x|A)\right],
\eeq
from $\v= 0$ where the determinant is given by \eq{GPY2}, to some
small value of $\v\ll 1$ (but such that $\v\D\gg 1$) where \eq{largesep2}
becomes valid. We shall parametrize this $\v$ as $\v= k/\D\ll 1$ with $k \gg 1$.
The result of the integration over $\v$ must be equal to the difference between the
right hand sides of \eqs{largesep2}{GPY2}. We write it as
\beq
\int_0^{\frac{k}{\D}} \!d\v\! \int\!d^4x\, \wp(x)
=  V \left[P\left(\!\frac{k}{\D}\!\right)-P(0)\right]
+2\pi \D\left[P''\left(\!\frac{k}{\D}\!\right)-P''(0)\right]
+ \log(k)+c-\frac{5}{3}\log(2\pi)-c_1-\frac{2}{3}\log \mu  +\O\left(\!\frac{k}{\D}\!\right).
\label{intwp}\eeq
Notice that $\log \D$ has cancelled in the difference in the r.h.s. We denote
\beq
c_2\equiv c-c_1-\frac{2}{3} \log \mu-\frac{5}{3}\log(2 \pi)\;.
\eeq
We know that the first two terms in \eq{intwp} come from far asymptotics.
Denoting by $\bar{\wp}$ our $\wp$ with subtracted asymptotic terms we have
\beq
\int_0^{\frac{k}{\D}} d\v \int\!d^4x\,\bar\wp(x)= \log  k+c_2+\O\left(\frac{1}{\D}\right)\;.
\la{intwp2}\eeq
In this integration we are in the domain $1/\v\gg 1$ and $\D\gg 1$ and we
can simplify the integrand dropping terms which are small in this domain.
At this point it will be convenient to restore temporarily the temperature dependence.
With $\beta\equiv 1/T$ our domain of interest is $1/\v\gg \beta$ and  $\D\gg \beta$.
Therefore we are in the small-$\beta$ domain and can expand $\bar\wp$
in series with respect to $\beta$:
\beq
\bar\wp= \frac{1}{\beta^2}\wp_0+\frac{1}{\beta}\wp_1+\O(\beta^0)\;.
\la{beta_exp}\eeq
As we shall see in a moment, only the first two terms are not small
in this domain and we need to know only them to compute $c_2$.
It is a great simplification because $\wp_{0,1}$ do not contain terms proportional
to $e^{-\bv r}$ since $\bv= 2\pi T-\v \to \infty$ at $\beta \to 0$, and what is left is time
independent. Moreover, what is left after we neglect exponentially small terms
are homogeneous functions of $r,s,\D,\v$ and we can turn to the dimensionless variables:
\beqa
\nn
&&\wp_0(r,s,\D,\v)= \frac{1}{\D}\widetilde{\wp}_0\left(\frac{r}{\D},
\frac{s}{\D},\frac{\v}{\D}\right),\\
\nn
&&\wp_1(r,s,\D,\v)= \frac{1}{\D^2}\widetilde{\wp}_1\left(\frac{r}{\D},
\frac{s}{\D},\frac{\v}{\D}\right)\;.
\eeqa
We rewrite the l.h.s. of \eq{intwp2} in terms of the new quantities:
\beq
\int\!d^4x\,\bar\wp(x)= \frac{\D^2}{\beta}\int\! d^3\tilde  x\,\widetilde{\wp}_0
+\D\int\! d^3\tilde x\,\widetilde{\wp}_1+\O(\beta)\;,
\la{intwp01}\eeq
where $\tilde x = x/\D$ is dimensionless. We see that it is indeed sufficient
to take just the first two terms in the expansion \ur{beta_exp} at $\beta\to 0$.
The integration measure can be written in terms of the dimensionless variables
$\tilde r = r/\D,\,\tilde s= s/\D$ as
\beq
d^3 \tilde x= 2\pi\, \tilde r d\tilde r\, \tilde s\, d\tilde s\;,
\eeq
where $\tilde r$ and $\tilde s$ are constrained by the
triangle inequalities $\tilde r\!+\!\tilde s\!<\!1$, $\tilde r\!+\!1\!<\!\tilde s$
and $\tilde s\!+\!1\!<\!\tilde r$, and we have integrated over the azimuth angle.

We have now to use the exact vacuum current to compute $\int\tilde\wp_{0,1}$.
First, it turns out that the first integral in \eq{intwp01} is zero.
This is good news because had it been nonzero, \eq{intwp2} could not be right as
its r.h.s. has no dependence on $\D$ other than possible $1/\D$ terms.
Second, we have noticed that the second integral in \eq{intwp01} is in fact
\beq
\int\!d^3\tilde x\, \widetilde{\wp}_1 = \frac{1}{\v \D+1}\,.
\la{miracle2}\eeq
Unfortunately, we were not able to verify it analytically
but we checked numerically that it holds with the precision of a few units
of $10^{-7}$ in the range of $\v\D$ between 0 and 15.
Combining \eqs{intwp01}{miracle2} we obtain for the l.h.s. of \eq{intwp2}
\beq
\nn
\int_0^{\frac{k}{\D}} d\v \int\!d^4x\,\bar\wp(x)=\D\int_0^{\!\frac{k}{\D}}
\frac{d\v}{\v \D+1}=\log(k+1)=\log k+\O\left(\frac{1}{k}\right).
\eeq
Therefore, we reproduce the r.h.s. of \eq{intwp2} and in addition find that $c_2= 0$.

\Eq{miracle2} is sufficient to extend the result for the determinant \ur{largesep2} valid
at $\v\D\gg 1$ to arbitrary values of $\v\D$, provided $\D\gg 1$ (the extension to
arbitrary values of $\bv\D$ is obtained by symmetry $\v\leftrightarrow \bv$). The final
result for the determinant to the $1/\D$ accuracy is
\beqa
\nn
\log\Det[\!-\!D^2]&=&V P(\v)+2\pi P''(\v)\,\D
+\left(1-\frac{4 \v}{3\pi}\right)\log\left(\v\D+1\right)
+\left(1-\frac{4 \bv}{3\pi}\right)\log\left(\bv\D+1\right)\\
&+&\frac{2}{3}\log(\mu\D)+c_1+\frac{5}{3}\log(2\pi)+\O\left(\frac{1}{\D}\right)
\la{final}\eeqa
where $\mu$ is the UV cutoff and the numerical constant $c_1$ is given by \eq{c1}.
This expression is finite at $\v\to 0,\,\bv\to 0$ and coincides with the GPY
result~\ur{GPY2} in these limits. At $\v\D\gg 1$ we restore the previous result, \eq{largesep2},
but now with the integration constant fixed: $c=\frac{2}{3}\log\mu+\frac{5}{3}\log(2\pi)+c_1$.
\Eq{final} is valid for any holonomy, i.e. for $\v,\bv\in[0,2\pi]$, and the only
restriction on its applicability is the condition that the dyon separation is large, $\D\gg 1$.

\subsection{$1/\D$ corrections}

\Eq{final} can be expanded in inverse powers of $\v\D,\bv\D$, which gives
$1/(\v\D),\;1/(\bv\D)$ (and higher) corrections; however, there are other $1/\D$
corrections which are not accompanied by the $1/\v,\;1/\bv$ factors: the aim of this
subsection is to find them using the exact vacuum current.

To this end, we again consider the case $\D\gg\frac{1}{\v},\,\frac{1}{\bv}$ such
that one can split the integration over $3d$ space into three regions shown in
Fig.~4. In the far-away region one can use the same vacuum current \ur{farCurrent}
as it has an exponential precision with respect to the distances to both dyons.
In the core regions, however, it is now insufficient to neglect completely the
field of the other dyon, as we did in section IV looking for the leading order.
Since we are now after the $1/\D$ corrections, we have to use the exact field
and the exact vacuum current of the caloron but we can neglect the exponentially
small terms in their separation.

Another modification with respect to section IV is that we find it more useful
this time to choose $\D$ as the parameter $\cal P$ in \eq{dvDet}.
We shall compute the $1/\D^2$ terms in $\d\Det(\!-\!D^2)/\d\D$ and then restore
the determinant itself since the limit of $\D\to\infty$ is already known.
Let us define how the KvBLL field depends on $\D$. As seen from \eq{APvB} the KvBLL field
is a function of $r,\;s,\;\v,\;\D$ only. We define the change in the separation
$\D\to \D+d\D$ as the symmetric displacement of each monopole center by $\pm d\D/2$.
It corresponds to
\beq
\frac{\d r}{\d \D}=\frac{\D^2+r^2-s^2}{4\D r},\qquad
\frac{\d s}{\d \D}=\frac{\D^2+s^2-r^2}{4\D s}.
\la{drsd}\eeq
We shall use the definition \ur{drsd} to compute the derivative of the caloron
field \ur{APvB} with respect to $\D$.

Let us start from the $M$-monopole core region. To get the $1/\D$ correction
to the determinant we need to compute its derivative in the $1/\D^2$ order
and expand correspondingly the caloron field and the vacuum current to this order.
Wherever the distance $r$ from the far-away L dyon appears in the equations, we
replace it by $r=(\D^2+2s\D\cos\theta+s^2)^{1/2}$ where $s$ is the distance from
the M-dyon and $\theta$ is the polar angle seen from the M-dyon center.
Expanding in inverse powers of $\D$ we get coefficients that are functions 
of $s,\cos\theta$. One can easily integrate over $\theta$ as the integration measure
in spherical coordinates is $2\pi s^2 ds\;d\cos\theta$. We leave out the 
intermediate equations and give only the end result for the integrand
in \eq{dvDet}. After integrating over $\cos\theta$ we obtain the following contribution
from the core region of the M monopole:
\beq
\left.\frac{\d\Det(\!-\!D^2)}{\d\D}\right|_{\rm M\;dyon\;core}
=-\frac{1}{\D^2}\int_0^R\!I_{1/\D^2}16\pi s^2 ds + {\cal  O}\left(\frac{1}{\D^3}\right),
\eeq
where $I_{1/\D^2}$ reads
\beqa
&&\nn I_{1/\D^2}=-\frac{\coth(s\v)}{12\pi^2 s^3}-\frac{\coth(s\v)}{9s}
-\frac{s\v^2\coth(s\v){{\rm csch}(s \v )}^2}{36}
-\frac{s^2\v^3\left(2+\cosh(2s\v)\right){{\rm csch}(s\v)}^4}{72}
\\
&&\nn+\frac{\v\left(-61+3\cosh(2s\v)+4\cosh(4s\v)\right)\coth(s\v)
{{\rm csch}(s\v)}^4}{96\pi s}
-\frac{\v^2\left(37+23\cosh(2s\v)+4\cosh(4s\v)\right)
\coth(s\v){{\rm csch}(s\v)}^4}{192\pi^2 s}
\\
&&\nn+\frac{s^2\v^4\left(4+\cosh(2s\v)\right){\coth(s\v)}^2
{{\rm csch}(s\v)}^4}{48\pi}
-\frac{s\v^3\left(54\cosh(s\v)+17\cosh(3s\v)+\cosh(5s\v)\right)  
{{\rm csch}(s\v)}^7}{384\pi}
\\
&&\nn+\frac{s\v^4\left(-406\cosh(s\v)-81\cosh(3s\v)+7\cosh(5s\v)\right)
{{\rm csch}(s\v)}^7}{2304 \pi^2}
-\frac{s^2\v^5\left(-24-33\cosh(2s\v)+\cosh(6s\v)\right)
{{\rm csch}(s\v)}^8}{1536\pi^2}
\\
&&\nn+\v\frac{6+ 13{{\rm csch}(s\v)}^2}{72}
+\frac{-10-5{{\rm csch}(s\v)}^2+9{{\rm csch}(s\v)}^4}{48\pi s^2}
+\v\frac{40+65{{\rm csch}(s\v)}^2+19{{\rm csch}(s\v)}^4}{192\pi^2 s^2}
\\
&&+\v^2\frac{-6-20{{\rm csch}(s\v)}^2+9{{\rm csch}(s\v)}^4
+27{{\rm csch}(s\v)}^6}{48\pi}
+\v^3\frac{ 24+98{{\rm csch}(s\v)}^2+285{{\rm csch}(s\v)}^4 
+234{{\rm csch}(s\v)}^6}{576\pi^2}\,.
\la{I12}\eeqa
Fortunately we are able to integrate this function analytically:
\beq
\int_0^R \!\!I_{1/\D^2}16\pi s^2 ds=\frac{1}{\v}-\frac{\pi^2+36\gamma_E+69}{27\pi}
+\frac{2\v(\v^2-3\pi\v+2\pi^2)}{9\pi}R^3+\frac{4(6\v\pi-2\pi^2-3\v^2)}{9\pi}R^2
+\frac{10(\v-\pi)}{3\pi}R-\frac{4\log( R\v/\pi)}{3\pi}.
\eeq
For the $L$ monopole core contribution one has to replace $\v$ by $\bv$.
Adding together contributions from L,M monopole cores we have
\beq
\left.\frac{\d\Det(\!-\!D^2)}{\d\D}\right|_{\rm cores}
=-\frac{1}{\D^2}\left[\frac{1}{\v}+\frac{1}{\bv}-2\frac{\pi^2+36\gamma_E+69}{27\pi}
-\frac{8}{9\pi}(3\v^2-6\pi\v+2\pi^2)R^2-\frac{4\log(R^2\v\bv/\pi^2)}{3\pi}\right].
\eeq

Now let us turn to the far-away region. Recalling \eq{J4asP} we
realize that the contribution of this region is determined by the
potential energy:
\beq\nn
\left.\frac{\d\Det(\!-\!D^2)}{\d\D}\right|_{\rm far}
=\int\!d^3x\, \d_{\D}P\left(\v+\frac{1}{r}-\frac{1}{s}\right)
= \frac{1}{2}P''(\v)\int\!d^3x \,\d_{\D}\left(\frac{1}{r}-\frac{1}{s}\right)^2
+\frac{1}{24}P^{\rm IV}(\v)\int\!d^3x \,\d_{\D}\left(\frac{1}{r}-\frac{1}{s}\right)^4.
\eeq
The integration region is the $3d$ volume with two balls of radius $R$ removed. We use
\beq
\int\!d^3x\,\d_{\D}\left(\frac{1}{r}-\frac{1}{s}\right)^2
=4\pi-\frac{16\pi R^2}{3\D^2},\qquad
\int\!d^3x\,\d_{\D}\left(\frac{1}{r}-\frac{1}{s}\right)^4
=\frac{2\pi}{3\D^2}\left[48\log\left(\frac{\D}{R}\right)-9\pi^2+8\right].
\eeq
Adding up all three contributions we see that the region separation radius $R$ gets
cancelled (as it should), and we get
\beq
\d_{\D}\log\Det(-D^2)=2\pi P''(\v)+\frac{1}{\D^2}\left[\frac{4}{3\pi}
\log\left(\frac{\v\bv\,\D^2}{\pi^2}\right)-\frac{1}{\v}
-\frac{1}{\bv}+\frac{50}{9\pi}+\frac{8\gamma_E}{3\pi}-\frac{23\pi}{54}\right]
\eeq
which can be easily integrated, with the result
\beq
\log\Det(-D^2)=2\pi P''(\v)\,\D+\frac{1}{\D}\left[\frac{1}{\v}+\frac{1}{\bv}
+\frac{23\pi}{54}-\frac{8\gamma_E}{3\pi}-\frac{74}{9\pi}
-\frac{4}{3\pi}\log\left(\frac{\v\bar\v\,\D^2}{\pi^2}\right)\right]
+\bar c+\O\left(\frac{1}{\D^2}\right)
\la{r12corr}\eeq
where $\bar c$ is the integration constant that does not depend on $\D$.
Comparing \eq{r12corr} with \eq{largesep1} at $\D\to\infty$ we conclude that
\beq
\bar c=V\,P(\v)+\frac{2}{3}\log\mu+\frac{3\pi-4\v}{3\pi}\log\v
+\frac{3\pi-4\bv}{3\pi}\log\bv+\frac{5}{3}\log(2\pi)+c_1
\eeq
and $c_1$ is given in \eq{c1}. 
We note that the leading correction, $\log\D/\D$, arises from the far-away region
and is related to the potential energy, similar to the leading $\D$ term. The terms
proportional to $\frac{1}{\v}$ and $\frac{1}{\bv}$ can be extracted from 
expanding \eq{final} (which is an additional independent check of \eq{miracle2}).
In fact, eqs.\ur{final} and \ur{r12corr} are complementary: \eq{final} sums up
all powers of $\frac{1}{\D\v},\,\frac{1}{\D\bv}$ but misses $\frac{\log\D}{\D}$
and $\frac{1}{\D}$ terms, whereas \eq{r12corr} collects all terms of that order
but misses higher powers of $\frac{1}{\D\v},\,\frac{1}{\D\bv}$.

\def\cM{{\cal{M}}}
\def\tr{{\rm tr}}

\section{Quantum weight of the KvBLL caloron}

\subsection{Quantum weight of a Euclidean pseudoparticle: generalities}

If a field configuration $\bar A_\mu$ is a solution of the Yang--Mills
Euclidean equation of motion, $D_\mu F_{\mu\nu}= 0$, its quantum weight
is the contribution of the saddle point to the partition function
\beq
{\cal Z}= \int DA_\mu \exp(-S[A]),\qquad
S[A]= \frac{1}{4g^2}\int d^4x\;F_{\mu\nu}^aF_{\mu\nu}^a.
\la{Z1}\eeq
The general field over which one integrates in \eq{Z1} can be written as
\beq
A_\mu= \bar A_\mu+a_\mu
\la{general_field}\eeq
where $\bar A_\mu$ is the classical solution corresponding to the local minimum of the
action and $a_\mu$ is the presumably small quantum oscillation about the solution.
One expands the action around the minimum,
\beq
S[A]= S[\bar A]-\frac{1}{g^2}\int\!d^4x\,a_\nu^a D^{ab}_\mu(\bar  A)F^b_{\mu\nu}(\bar A)
+\frac{1}{2g^2}\int\!d^4x\,a_\mu^a\,W^{ab}_{\mu\nu}(\bar  A)\,a_\nu^b+\O(a^3),
\la{S_expan}\eeq
where the linear term is in fact absent since $\bar A$ satisfies the
equation of motion, and the quadratic form is
\beqa
\la{W1}
W_{\mu\nu}^{ab}(\bar A)&= &-D^2(\bar A)^{ab}\delta_{\mu\nu}+(D_\mu  D_\nu)^{ab}(\bar A)
-2f^{acb}F^c_{\mu\nu}(\bar A),\\
D_\mu^{ab}(\bar A)&= &\partial_\mu\delta^{ab}+f^{acb}\bar A_\mu^c.
\la{D1}\eeqa
We have written the covariant derivative in the adjoint representation; the relation
with the fundamental representation is given by  $a_\mu= -ia_\mu^at^a,\;
\Tr(t^at^b)= \half\delta^{ab}$ and similarly for $F_{\mu\nu}$, etc.
The 1-loop approximation to the  quantum weight corresponds to evaluating
\eq{Z1} in the Gaussian approximation in  $a_\mu$, hence $\O(a^3)$ terms in
\eq{S_expan} have been neglected.

The quadratic form \ur{W1} is highly degenerate since any fluctuation of  the type
$a_\mu^a= D_\mu^{ab}(\bar A)\Lambda^b(x)$ corresponding to an infinitesimal gauge transformation
of the saddle-point field $\bar A$, nullifies it. Therefore, one has to impose a gauge-fixing
condition on $a_\mu$. The conventional choice is the background Lorenz gauge
$D_\mu^{ab}(\bar A)a_\mu^b= 0$: with this condition imposed the  operator $W$ simplifies
as the second term in \eq{W1} can be dropped. Fixing this gauge,  however, brings in the
Faddeev--Popov ghost determinant $\Det(-D^2(\bar A))$.

To define the path integral, one decomposes the fluctuation field in the
complete set of the eigenfunctions of the quadratic form,
\beq
a_\mu^a(x)= \sum_nc_n\psi^a_{\mu\,n}(x),\qquad
W^{ab}_{\mu\nu}\psi^b_{\nu\,n}= \lambda_n\psi^b_{\mu\,n},\qquad
D^{ab}_\mu\psi^b_{\mu\,n}= 0,
\la{decomp}\eeq
and implies that the path integral is understood
as the integral over Fourier coefficients in the decomposition:
\beq
DA_\mu(x) =  \prod_n \frac{dc_n}{\sqrt{2\pi}},
\la{measure_def}\eeq
The quadratic form \ur{W1} has a finite number
of zero modes related to the moduli space of the solution. Let the  number of zero modes be $p$
(for a self-dual solution with topological charge one $p= 4N$ for the  $SU(N)$ gauge group~\cite{CWS}).
Let $\xi_i,\,i= 1...p$, be the set of collective coordinates  characterizing the classical solution,
of which the action $S[\bar A]$ is in fact independent. The zero modes  are
\beq
\psi_{\mu\,i}^a(x)= \frac{\d \bar A_\mu^a(x,\xi)}{\d  \xi_i}-D^{ab}_\mu(\bar A)\Lambda^b_i(x)
\la{zero_modes}\eeq
where the second term is subtracted in order for the zero modes to satisfy the background
Lorenz condition, $D_\mu^{ab}\psi^b_{\mu\,i}=0$. The $p\times p$ metric tensor
\beq
g_{ij}= \int\!d^4x\,\psi_{\mu\,i}^a\psi_{\mu\,j}^a
\la{metric1}\eeq
defines the metric of the moduli space. Its determinant is actually the  Jacobian for
passing from integration over zero-mode Fourier coefficients  $c_i,\,i= 1...p$, in \eq{measure_def}
to the integration over the collective coordinates $\xi_i,\,i= 1...p$:
\beq
\prod_{i= 1}^p\frac{dc_i}{\sqrt{2\pi}}= J\prod_{i= 1}^p  d\xi_i\left(\frac{1}{\sqrt{2\pi}}\right)^p,
\qquad J=\sqrt{\det g_{ij}}.
\la{Jacobian}\eeq

Finally, one has to normalize and regularize the ghost determinant  $\Det(-D^2)$ and the Gaussian
integral of the quadratic form. One usually normalizes the contribution  of a pseudoparticle
to the partition function by dividing it by the free (i.e. zero  background field) determinants,
and regularizes it by dividing by the determinants of the $-D^2$ and  $W_{\mu\nu}$ operators
shifted by the Pauli--Villars mass $\mu$~\cite{tHooft,Bernard}. It means  that $\Det(-D^2)$ is replaced
by the `quadrupole' combination
\beq
\Det(-D^2)_{\rm  n,\,r}= \frac{\Det(-D^2)}{\Det(-\d^2)}
\frac{\Det(-\d^2+\mu^2)}{\Det(-D^2+\mu^2)}
\la{quadrupole1}\eeq
and similarly for the determinant of the quadratic form,
\beq
\Det'(W_{\mu\nu})_{\rm  n,\,r}= \frac{\Det'(W_{\mu\nu})}{\Det(-\d^2\delta_{\mu\nu})}
\frac{\Det(-\d^2\delta_{\mu\nu}+\mu^2)}{\Det'(W_{\mu\nu}+\mu^2)},
\la{quadrupole2}\eeq
where the prime indicates that only the product of nonzero eigenvalues  is taken.
In the integration over Pauli--Villars fields, the zero eigenvalues are  shifted
by $\mu^2$. Hence the integration over the zero-mode Fourier  coefficients in the
Pauli--Villars fields produces the factor
\beq
\prod_{i= 1}^p\int\!\frac{dc_i}{\sqrt{2\pi}}\,\exp\left[-\frac{1}{2g^2}c_i^2(0+\mu^2)\right]
= \left(\frac{g}{\mu}\right)^p
\la{PVzm}\eeq
which has to be taken in the minus first power. Finally, one obtains the  following
normalized and regularized expression for the 1-loop quantum weight of a  Euclidean pseudoparticle:
\beq
{\cal Z}= \int \prod_{i= 1}^p d\xi_i\,e^{-S[\bar  A]}\left(\frac{\mu}{g\sqrt{2\pi}}\right)^p\,
J\,\left(\Det'(W_{\mu\nu})_{\rm n,\,  r}\right)^{-\frac{1}{2}}\,\Det(-D^2)_{\rm n,\,r}.
\la{Z2}\eeq
If the saddle-point field $\bar A_\mu$ is (anti)self-dual there is a  remarkable relation between
the two determinants~\cite{BC}: $\Det'(W_{\mu\nu})_{\rm n,\,r}=\Det^4(-D^2)_{\rm n,\, r}$
which is satisfied if the background field is decaying fast enough at  infinity and
the Hilbert space of the eigenfunctions of the two operators is well  defined. This is
the case of the KvBLL caloron but not the case of a single BPS dyon  having a Coulomb
asymptotics. To define the dyon weight properly, one would need to  consider it in a spherical box,
which would violate most of the statements in this subsection. For this  reason we prefer to
consider the well-defined quantum weight of the KvBLL caloron in which  case the product of two
determinants in \eq{Z2} becomes just $\Det^{-1}(-D^2)$.

\subsection{KvBLL caloron moduli space}

The KvBLL moduli space has been studied in the original  papers~\cite{KvB,LL};
in particular in ref.~\cite{KvB} the metric tensor $g_{ij}$ \ur{metric1}  has been
explicitly computed. We briefly review these results and adjust them to  our needs.

The KvBLL classical solution has 8 parameters for the $SU(2)$ gauge  group.
These are the four center-of mass positions $z_\mu$ and the four  quaternionic variables
$\zeta= \rho U $  corresponding to the constituent monopoles relative  position in space
and one global gauge transformation, see Appendix A.3. The moduli space  of the KvBLL caloron
is a product of the base manifold $\re^3 \times S^1$ parameterized by  the $\vec z\in \re^3$
and $z_4\in [0,1]$, and the non-trivial part of the moduli space  parameterized by the
quaternion $\zeta$. It should be noted that the change $\zeta\to -\zeta$, corresponding to the
center of the $SU(2)$, leaves $\bar A_\mu(x)$ invariant, such that one  has to mod-out
this symmetry.

The 8 zero modes $\psi^a_{\mu\,i}$ \ur{zero_modes} satisfying the background
Lorenz condition have been explicitly found in ref.~\cite{KvB}. If one parametrizes
the unitary matrix through Euler angles,
\beqa\la{U}
U= e^{-i\Upsilon\frac{\tau_3}{2}}e^{i(\frac{\pi}{2}-\theta)\frac{\tau_2}{2}}
e^{-i\varphi\frac{\tau_1}{2}},\qquad
&0\leq\Upsilon\leq 4\pi,\quad 0\leq\varphi\leq2\pi,\quad  0\leq\theta\leq\pi,
\eeqa
the metric is~\cite{KvB}
\beq
d s^2= (2\pi)^2[2dz_\mu dz_\mu+(1+8\pi^2\omega\bar\omega\rho^2)
\left(4d\rho^2+\rho^2  d^2\Omega\right)+\rho^2(1+8\pi^2\omega\bar\omega\rho^2)^{-1}d\Sigma^2_3]
\label{mainmetric}\eeq
where
\beq
d^2\Omega= \sin^2\!\theta\, d\varphi^2+d\theta^2,\qquad
d\Sigma_3= d\Upsilon+\cos\theta\, d\varphi.
\la{Omega_sigma}\eeq
The first part describes the flat metric of the base manifold $\re^3  \times S^1$,
the remainder forms the non-trivial part of the metric. The variables  are inside the ranges
$\rho \in[0,\infty),\;\theta \in [0,\pi),\;\phi \in [0,2\pi),\; \Upsilon  \in [0,4 \pi)/Z_2
=  [0,2\pi)$ for the non-trivial part, and $z_4 \in [0,1], \;z_i\in \re  $ for translational modes.

The collective coordinate Jacobian is immediately found from  \eq{mainmetric}:
\beq
J= \sqrt{\det(g_{ij})}   =
8\,(2\pi)^8\,\rho^3\,(1+8\pi^2\rho^2\omega\bar\omega)\,\sin\theta\,.
\la{Jac1}\eeq
The factor $\sin\theta$ is needed to organize the orientation $SO(3)$  Haar measure normalized
to unity,
\beq
\int\!d^3\O = \frac{1}{8\pi^2}\int_0^{2\pi}\!d\Upsilon\!\int_0^{2\pi}\!d\varphi\!
\int_0^\pi\!d\theta\,\sin\theta = 1,
\la{Haar}\eeq
and the KvBLL measure written in terms of the caloron center, size and  orientation
becomes
\beq
\int\!d^3z\!\int\!dz_4\!\int\!d^3\O\!\int\!d\rho\,\rho^3\,
(1+8\pi^2\omega\bar\omega\rho^2)\,16\,(2\pi)^{10}.
\la{meas1}\eeq
This must be multiplied by the factors  $\left(\frac{\mu}{g\sqrt{2\pi}}\right)^8$ and
$\exp(-S[\bar A])= \exp\left(-\frac{8\pi^2}{g^2}\right)$ according to  \eq{Z2}.
As the result, the KvBLL measure is
\beq
\int\!d^3z\!\int\!dz_4\!\int\!d^3\O\!\int\!\frac{d\rho}{\rho^5}\,(1+8\pi^2\omega\bar\omega\rho^2)\,
(\mu\rho)^8\frac{1}{4\pi^2}\left(\frac{8\pi^2}{g^2}\right)^4\,e^{-\frac{8\pi^2}{g^2}}.
\la{meas2}\eeq
When the holonomy is trivial ($\omega\!=\!0$ or $\bar\omega\!=\!0$) it becomes the well-known
measure of the BPST instanton~\cite{Bernard} or that of the  Harrington--Shepard caloron~\cite{GPY}.
The difference between the two is that in the first case one integrates over any $z_4$
whereas in the second case the $z_4$ integration is restricted to $z_4\in [0,\frac{1}{T}]$.
\Eq{meas2} would have been the full result in the ${\cal N}\!=\!1$ supersymmetric theory
where the determinant over nonzero modes is cancelled by the gluino determinant. In that case
one would need to add the integral over Grassmann variables corresponding to the gluino
zero modes.

\subsection{Combining the Jacobian and the determinant over nonzero  modes}

According to the general \eq{Z2}, we have now to multiply \eq{meas2} by the
(regularized and normalized) determinant over nonzero modes, which has been
calculated in \eq{final}. First of all, we notice that $\Det^{-1}(-D^2)$
brings in an additional UV divergent factor $\mu^{-\frac{2}{3}}$. In combination
with the classical action and the factor $\mu^8$ coming from zero modes, it
produces
\beq
\mu^{\frac{22}{3}}\,e^{-\frac{8\pi^2}{g^2(\mu)}}=\Lambda^{\frac{22}{3}}
\la{transmutation}\eeq
where $\Lambda$ is the scale parameter obtained here through the
`transmutation of dimensions'.

We notice further that $\Det^{-1}(-D^2)$ is independent of the $SU(2)$ orientation
$\O$ and of $z_4$. Therefore, we integrate over these variables, which gives unity.
Next, we introduce the centers of the constituent BPS dyons $\vec z_{1,2}$ such that
$|z_1-z_2|=\D=\pi\rho^2$ and write
\beq
\int\!d^3 \vec z_1\, d^3 \vec z_2= \int\!d^3\!\!\left(\frac{\vec z_1+\vec z_2}{2}\right)
d^3(\vec z_1- \vec z_2)=4\pi \int\!d^3z\, \D^2 d\D
= 8\pi^{\frac{3}{2}}\int\!d^3z\,d\rho\,\D^{\frac{5}{2}}.
\eeq
Therefore, integration over $d^3zd\rho$ in \eq{meas2} can be traded for integrating
over the dyon positions in space, $\vec z_{1,2}$. Lastly, we restore the temperature
from dimensional considerations and obtain our final result for the 1-loop quantum
weight of the KvBLL caloron, written in terms of the coordinates of the dyon centers:
\beqa\nn
{\cal Z}_{\rm KvBLL}&=&\int d^3z_1 \, d^3z_2\, T^6\,C\left(\frac{8\pi^2}{g^2}\right)^4\!
\left(\frac{\Lambda e^{\gamma_E}}{4\pi T}\right)^{\frac{22}{3}}
\left(\frac{1}{T\D}\right)^{\frac{5}{3}}
\left(2\pi+\frac{\v \bv }{T}\D\right)
\left(\v \D+1\right)^{\frac{4 \v}{3\pi T}-1}\left(\bv\D+1\right)^{\frac{4 \bv}{3\pi T}-1}\\
\nn\\
\la{Zfull}
&\times &\exp\left[-V\, P(\v)-2\pi\,\D\, P''(\v)\right],
\eeqa
where
\beq
C=\frac{64}{\pi^2}\,\exp\left[{\frac{8}{9}-\frac{16\,{\gamma_E}}{3}
+\frac{2\pi^2}{27}+\frac{4\,{\zeta}'(2)}{\pi^2}}\right]= 1.031419972084
\la{C}\eeq
and $P(\v)$ is the potential energy
\beq
P(\v)=\frac{1}{12\pi^2T}\v^2\bv^2,\qquad
P''(\v)=\frac{1}{\pi^2 T}\left[\pi  T\left(1-\frac{1}{\sqrt{3}}\right)-\v\right]
\left[\bv-\pi T\left(1-\frac{1}{\sqrt{3}}\right)\right],\qquad \bv=2\pi T-\v.
\la{PandPpp}\eeq

We have collected the factor $4\pi e^{-\gamma_E}T/\Lambda$ because it is the natural
argument of the running coupling constant at nonzero  temperatures~\cite{coupling,DO}.
Here $\Lambda$ is the scale parameter in the Pauli--Villars regularization scheme that
we have used. It is related to scale parameters in other schemes:
$\Lambda_{\rm PV}=e^{\frac{1}{22}}
\Lambda_{\overline{\rm MS}}=40.66\cdot\exp\left(-\frac{3\pi^2}{11 N^2}\right)\,
\Lambda_{\rm lat}$~\cite{Has}. The factor $g^{-8}$ is not renormalized
at the one-loop level: it starts to `run' at the 2-loop level, see below. \\

The KvBLL caloron weight \ur{Zfull} has been derived assuming the separation between
constituent dyons is large in temperature units ($\D\gg \frac{1}{T}$) 
but the holonomy is arbitrary: $\half\Tr\,L\in[-1,1]$ corresponding to
$\v,\bv \in [0,2\pi T]$. It means that \eq{Zfull} is valid not only for 
well-separated but also for overlapping dyons. 

\subsection{The limit of large dyon separation}

In the limit when the separation of dyons is larger than their core sizes,
$\D\gg \frac{1}{\v},\frac{1}{\bv}$, the caloron weight simplifies to
\beqa
\nn
{\cal Z}_{\rm KvBLL}&=&\exp\left[-VT^3\,\frac{4\pi^2}{3}\nu^2(1-\nu)^2\right]
\int\!d^3z_1\, d^3z_2\, T^6\,(2\pi)^{\frac{8}{3}}\,C\left(\frac{8\pi^2}{g^2}\right)^4
\left(\frac{\Lambda e^{\gamma_E}}{4\pi T}\right)^{\frac{22}{3}}
\nu^{\frac{8}{3}\nu}\,(1-\nu)^{\frac{8}{3}(1-\nu)}\\
&\times &\exp\left[-2\pi\,\D\,T\,\left(\frac{2}{3}-4\:\nu(1-\nu)\right)\right]
\la{Zfar}\eeqa
where we have introduced the dimensionless quantity $\nu=\frac{\v}{2\pi T}\in[0,1]$.
In subsection V.C we have calculated the $\frac{1}{\D T}$ correction to the determinant,
see \eq{r12corr}. Another correction arises from the Jacobian \ur{Jac1} which cancels
the $\frac{1}{\v},\,\frac{1}{\bv}$ terms in \eq{r12corr}. As a result, we get the following
correction factor to \eq{Zfar}
\beq
\exp\left[\frac{1}{\D T}\left(\frac{4}{3\pi}\log\left[\nu(1-\nu)(2\D T)^2\right]
+c_{1/\D}\right)+\O\left(\frac{1}{(\D T)^2}\right)\right],\qquad
c_{1/\D}=\frac{74}{9\pi}+\frac{8\gamma_E}{3\pi}-\frac{23\pi}{54}=1.946\,.
\la{Zfarcorr}\eeq

One can define the interaction potential between $L,M$ dyons as
\beq
V_{\rm LM}(\D)=\D\,T^2\,2\pi\left(\frac{2}{3}-4\:\nu(1-\nu)\right)
-\frac{1}{\D}\left(\frac{4}{3\pi}\log\left[\nu(1-\nu)(2\D T)^2\right]
+c_{1/\D}\right)+\O\left(\frac{1}{\D^2 T}\right).
\la{LMinteract}\eeq
This interaction is a purely quantum effect: classically $L,M$ dyons do not
interact at all as the KvBLL caloron of which they are constituents has the
same classical action for all $L,M$ separations. Curiously, the interaction 
potential has the familiar ``linear $+$ Coulomb" form. Both terms depend
seriously on the holonomy: the Polyakov line at spatial infinity 
is $\half\Tr\,L=\cos(\pi\nu)$. In the range 
$0.787597\!<\!\half|\Tr\,L|\!<\!1$ dyons experience asymptotically 
a constant attraction force; in the complementary range 
$\half|\Tr\,L|\!<\!0.787597$ it is repulsive. It should be noted
that in its domain of applicability $\D\gg \frac{1}{\v},\frac{1}{\bv}\geq \frac{1}{2\pi T}$,
the second term in \eq{LMinteract} is a small correction as compared 
to the linear rising (or linear falling) interaction.

\subsection{2-loop improvement of the result}

The factor $g(\mu)^{-8}$ in \eq{Zfull} is the bare coupling which is  renormalized
only at the 2-loop level. In the case of the zero-temperature instanton  one can
unambiguously determine the 2-loop instanton weight without explicit  2-loop calculations
from the requirement that it should be invariant under the simultaneous  change
of the UV cutoff and of the bare coupling given at that cutoff, such  that the scale parameter
\beq
\Lambda=\mu\,\exp\left(-\frac{8\pi^2}{b_1g^2(\mu)}\right)\,
\left(\frac{16\pi^2}{b_1g^2(\mu)}\right)^{\frac{b_2}{2b_1^2}}\,
\left[1+O\left(g^2(\mu)\right)\right],\qquad
b_1=\frac{11}{3}N,\qquad
b_2=\frac{34}{3}N^2,
\eeq
remains invariant. The result~\cite{DP84} is that one has to replace the  combination of the
bare coupling constants
\beq
\left(\frac{8\pi^2}{g^2(\mu)}\right)^{2N}\!\exp\left(-\frac{8\pi^2}{g^2(\mu)}\right)
\;\longrightarrow\;
\beta(\tau)^{2N}\,\exp\left[-\beta^{\rm II}(\tau)
\!+\!\left(2N\!-\!\frac{b_2}{2b_1}\right)\!\frac{b_2}{2b_1}
\frac{\ln\beta(\tau)}{\beta(\tau)}+\O\left(\frac{1}{\beta(\tau)}\right)\right]
\la{2loop}\eeq
where
\beq
\beta(\tau)=b_1\ln\frac{\tau}{\Lambda},\qquad
\beta^{\rm II}(\tau)=\beta(\tau)+\frac{b_2}{2b_1}\ln\frac{2\beta(\tau)}{b_1},\\
\la{beta2}\eeq
and $\tau$ is the scale of the pseudoparticle, which is $1/\rho$ in the  instanton case.
In the case of the KvBLL caloron with widely separated constituents one  has to take the
temperature scale, $\tau=4\pi T e^{-\gamma_E}$. Thus, the 2-loop  recipe is to replace
the factor $(8\pi^2/g^2)^4\,(\Lambda e^{\gamma_E}/4\pi T)^{22/3}$ in  \eqs{Zfull}{Zfar}
by the r.h.s. of \eq{2loop}.

In contrast to the zero-temperature instanton, in the KvBLL caloron case
this replacement is not the only effect of two loops. In particular, the  potential energy
$P(\v)$ is modified in 2 loops~\cite{Belyaev}. Nevertheless, the above  modification is
a very important effect of two loops, which needs to be taken into  account if one wants
to make a realistic estimate of the density of calorons with non-trivial  holonomy at a
given temperature. We remark that the additional large factor $4\pi e^{-\gamma_E}\approx 7.05551$
makes the running coupling numerically small even at $T\simeq \Lambda$ ($1/\beta(\tau)\simeq 0.07$),
which may justify the use of semiclassical methods at temperatures around the phase transition.
This numerically large scale is not accidental but originates from the fact that it is
the Matsubara frequency $2\pi T$ rather that $T$ itself which serves as a scale in all
temperature-related problems. The additional order-of-unity factor $2e^{-\gamma_E}$
is specific for the Pauli--Villars regularization scheme used.

\section{Caloron density and instability of the trivial holonomy}

Since the caloron field has a constant $A_4$ component at spatial  infinity,
it is strongly suppressed by the potential energy $P(\v)$, unless  $\v= 0,2\pi T$
corresponding to trivial holonomy. Nevertheless, one may ask if the free  energy
of an ensemble of calorons can override this perturbative potential. We  make
below a crude estimate of the free energy of non-interacting KvBLL  calorons.
We shall consider only the case of small 
$\v<\pi T\left(1\!-\!\frac{1}{\sqrt{3}}\right)$.
If $\v$ exceeds this value the integral over dyon separations in  \eq{Zfull}
diverges, meaning that calorons with holonomy far from trivial  ``ionize''
into separate dyons. We shall not consider this case here but restrict  ourselves
to small values of $\v$ where the integral over the separation between  dyon
constituents converges, such that one can assume that KvBLL calorons are
in the ``atomic'' phase. Integrating over the separation $\D$ in  \eq{Zfull}
gives the ``fugacity'' of calorons:
\beqa
\la{zeta1}
\zeta&= &T^3\, f\left(T/\Lambda\right)\,I(\nu),\\
\la{fTL}
f\left(T/\Lambda\right)&= &8\pi^2\,C\,\beta^4\,
\exp\left[-\beta^{\rm  II}\!+\!\left(4\!-\!\frac{34}{11}\right)\!
\frac{34}{11}\frac{\ln\beta}{\beta}\right],
\qquad \beta= \frac{22}{3}\ln\frac{4\pi T}{\Lambda e^{\gamma_E}},
\qquad \beta^{\rm II}= \beta+\frac{34}{11}\ln\frac{3\beta}{11},\\
\la{Inu}
I(\nu)&= &\int_0^\infty\!\!dR\,R^{\frac{1}{3}}\left[1+2\pi\nu(1\!-\!\nu)R\right]
(2\pi\nu R+1)^{\frac{8}{3}\nu-1}(2\pi(1\!-\!\nu)  R+1)^{\frac{8}{3}(1\!-\!\nu)-1}
\exp\left[\!-\!2\pi R\left(\frac{2}{3}\!-\!4\nu\!+\!4\nu^2\right)\right]
\eeqa
where we have introduced the dimensionless separation, $R=\D T$, and
the dimensionless $\nu=\frac{\v}{2\pi T}$. One should be cautioned that \eq{Zfull}
has been derived for $R\gg 1$, therefore the caloron fugacity is evaluated
accurately if the integral \ur{Inu} is saturated in the large-$R$ region.

Assuming the Yang--Mills partition function is governed by a non-interacting
gas of $N_+$ calorons and $N_-$ anticalorons, one writes their grand 
canonical partition function as
\beq
{\cal Z}_{\rm  cal}= \exp\left[-VT^3\,\frac{4\pi^2}{3}\nu^2(1-\nu)^2\right]
\sum_{N_+,N_-}\frac{1}{N_+!N_-!}\left(\!\int\!d^3z\,\zeta\right)^{N_++N_-}
= \exp\left[-VT^3 {\cal F}(\nu,T)\right],
\la{Zgas}\eeq
where ${\cal F}(\nu,T)$ is the free energy of the caloron gas, including  
the perturbative potential energy:
\beq
{\cal F}(\nu,T)=
\frac{4\pi^2}{3}\nu^2(1-\nu)^2-2f(T/\Lambda)\,I(\nu).
\la{Fgas}\eeq
We plot the free energy as function of $\nu$ in Fig.~5 at several temperatures.
The function $f(T/\Lambda)$ rapidly drops with increasing temperature. Therefore,
at high temperatures the perturbative potential energy prevails, and the minimal
free energy corresponds to trivial holonomy, However, at  $T\approx\Lambda$ the
caloron fugacity becomes sizable, and an opposite trend is observed. In this model,
$T_c= 1.125\,\Lambda$ is the critical temperature where the trivial holonomy becomes
an unstable point, and the system rolls towards large values of $\v$ where the
present approach fails since at large $\v$ calorons anyhow have to ``ionize''
into separate dyons.

\begin{figure}[t]
\centerline{
\epsfxsize=0.4\textwidth
\epsfbox{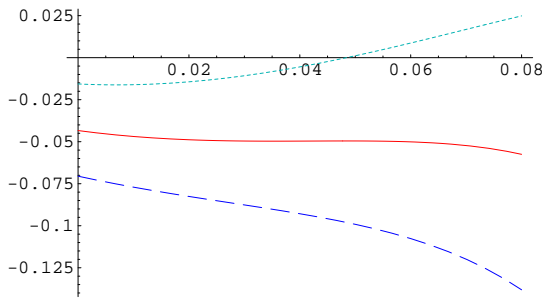}}
\caption{\label{fig:instability} Free energy of the caloron gas in units of $T^3V$ at
$T= 1.3\Lambda$ (dotted), $T= 1.125\Lambda$ (solid) and  $T= 1.05\Lambda$ (dashed)
as function of the asymptotic value of $A_4$ in units of $2\pi T$.}
\end{figure}

Although several simplifying assumptions have been made in this derivation,
it may indicate the instability of the trivial holonomy at temperatures
below some critical one related to $\Lambda$.

\section*{Acknowledgements}
N.G. and S.S. thank the foundation for non-commercial programs
`Dynasty' for partial support. N.G. is grateful to NORDITA where
part of the present work has been done, for hospitality.

\appendix
\section{ADHM construction for the BPS dyon and the KvBLL caloron}

\subsection{General ADHM construction}

The basic object in the ADHM construction~ \cite{ADHM} is the
$k\times (k+1)$ quaternionic-valued matrix $\Delta$ which is taken
to be linear in the space-time variable $x$:
\beq
\Delta(x)= \cA+\cB x,\qquad x\equiv x_\mu \sigma_\mu,\qquad
\sigma_\mu= (i\vec\tau,1_2) \label{cAcB} \;.
\eeq
The ADHM gauge potential is given by
\beq
A_\mu(x)= v^\dagger(x)\partial_\mu v(x),
\label{eq:AADHM}\eeq
where $v(x)$ is a $(k+1)$ dimensional
quaternionic vector, the normalized solution to
\beq
\Delta^\dagger(x)v(x)= 0,
\label{deltav}\eeq
and $k$ is the
topological charge of the gauge field. An important property of
the ADHM construction is that the operator
$\Delta^{\dagger}(x)\Delta(x)$ is a real-valued matrix:
\beq
f= (\Delta(x)^\dagger \Delta(x))^{-1}\in\re^{k \times
k}\label{deff} \;. \eeq In what follows we shall use the equation
\beq \Delta f\Delta^\dag= 1-vv^\dag \label{vv2f} \;.
\eeq
It becomes obvious when one notes that both sides are projectors onto
the space orthogonal to the vector $v$, which follows from $v^\dag
v= 1$, $\Delta^\dag v= 0$.

In the case of finite temperatures, because of the infinite number
of copies of space in the compact direction, the topological
charge $k= \infty$, and it is convenient to make a discrete Fourier
transformation with respect to the infinite range indices. The
Fourier transformed $v(x)$ are $2\times 2$ matrix-valued functions
$v(x_\mu, z)$ of a new variable $z\in [-1/2,1/2]$ and $\Delta$
becomes a differential operator in $z$.

\subsection{ADHM construction for the BPS dyon}

As stated above, at nonzero temperatures the essence of the ADHM
construction is the introduction of $2\times 2$ matrix-valued
functions $v(x_\mu, z)$. The scalar product is defined as
\beq
\langle v_1|v_2\rangle= \int^{1/2}_{-1/2} v^+_1(x_\mu, z)v_2(x_\mu,  z)dz \;.
\eeq
For the BPS dyon solution $v$ has been found by
Nahm~\cite{Nahm80}: \beq v(x_\mu, z)= \sqrt{\frac{\v r}{\sinh(\v
r)}}\exp( iz\v x^\dag) \la{w_monopole}\eeq where
$\sigma^\dag_{\mu}= (1_2,-i\vec{\tau}),\;x^\dag= x_\mu\sigma^\dag_\mu$
and $r= |\vec{x}|$). The matrix-valued function $v$ is the solution 
of the equation
\beq
\Delta^\dag(x)v(x,z)= 0,\;\;\;\;\;\Delta^\dag(x)= i\partial_z+\v
x^\dag \la{p2}\eeq normalized to unity, \beq \langle v|v\rangle= 1.
\eeq The gauge field is expressed through $v$ as \beq
A_\mu= \langle v|\partial_\mu v\rangle. \la{Amu}\eeq We use
anti-hermitian fields such that the covariant derivative is
$D_\mu= \d_\mu+A_\mu$. Comparing \eq{p2} with the general \eq{cAcB}
we conclude that in this case
\beq \cA= -i\d_z,\;\;\;\;\;\cB= \v.
\la{ABmono}
\eeq

\Eq{w_monopole} corresponds to the `hedgehog' gauge. However we
find it more convenient to work in the `stringy' gauge where
$A_\mu$ has a pure gauge string-like singularity. One proceeds
from the `hedgehog' gauge to the `stringy' gauge using the
singular gauge transformation (see e.g.~\cite{DPSUSY})
\beq
v\rightarrow v^s= v S_-^\dag, \quad A_\mu \rightarrow A_\mu^s= S_-
A_\mu S_-^\dag + S_-\partial_\mu S^\dag_-
\eeq
with
\beq 
S_-=e^{-i\frac{\phi}{2}\tau^3}e^{i\frac{\pi-\theta}{2}\tau_2}
e^{-i\frac{\phi}{2}\tau^3} \label{Sm}
\eeq
having the property that it ``gauge-combs" $A_4$ at spatial
infinity to a fixed (third) direction:
\beq
S_- n^a\tau^a S_-^\dag=\tau^3 \;.
\eeq
In the `stringy' gauge
\beq
v^s=S_{-}^\dag\sqrt{\frac{\v r}{\sinh (\v r)}}\exp[z\v(i x_4+r\tau^3)]\;.
\label{vmonostr}\eeq
One can check that
$A_\mu= \langle v^s|\d_\mu v^s\rangle$ gives the M dyon field in the
`stringy' gauge as in \eq{Mdyon}. We note that in the `stringy' gauge 
$v^s$ has a remarkable property
\beq
v^s(x_4+n,\vec x)= e^{i n\v z}v^s(x_4,\vec x)\;.
\label{BPSpln}\eeq

\subsection{ADHM construction for the KvBLL caloron \label{ADHM}}

Unfortunately, the original paper \cite{KvB} does not present an
explicit expression for $v$, the main ingredient of the ADHM
construction. We could have used ref.~\cite{LL} but it seems that
ref.~\cite{KvB} is more informative in some other respects.
Therefore, we have to calculate $v$ ourselves.

From the point of view of the original ADHM construction $v$ is a
quaternionic vector of infinite length since finite-temperature
field configuration can be viewed as an infinite set of equal
strips, the total topological charge in $\re^4$ being infinite.
The bracket is formally defined as a contraction along this
infinite-dimension side:
\beq
\langle v|\tilde v\rangle\equiv
v^\dag \tilde v \;.
\eeq
The gauge potential results from
\beq
A_\mu(x)= v^\dagger(x)\partial_\mu
v(x),\;\;\;\;\;D_\mu= \d_\mu+A_\mu\label{eq:AADHM1} \;.
\eeq
The vector $v(x)$ is the normalized solution of the equation
\beq
\Delta^\dagger(x)v(x)= 0,\;\;\;\;\;
\Delta(x)= \left(\!\!\bea{c}\lambda\\B-x\eea\!\!\right),
\la{NahmEq}\eeq
where $B$ is a square quaternionic matrix, $\lambda$ is an
(infinite) quaternionic vector, $x\equiv x_\mu\sigma_\mu,\;
\sigma_\mu=(i\vec\tau,1_2)$. Introducing the notations
\beq
v(x)=\Mphi^{-\half}(x)\left(\!\!\bea{c}-1\\u(x)\eea\!\!\right),\quad
u(x)=(B^\dagger-x^\dagger)^{-1}\lambda^\dagger \;,
\label{eq:vu}\eeq
\eq{NahmEq} becomes
\beq
(B^\dag-x^\dag)u(x)= \lambda^\dag .
\eeq
The inverse of the matrix $(B^\dag-x^\dag)$ exists almost in all points.
The points where it does not exists, are monopole positions. We are interested in
those singular points that lie in the interval $0<x_4<1$ (we have
rescaled the units to set temperature $T= 1$).
The unknown function $\Mphi(x)$ is determined from the
normalization of $v$:
\beq
\Mphi(x)= 1+u^\dag u  \;.
\eeq

The formalism of infinite-dimensional matrices is not convenient.
Following Nahm~\cite{Nahm80} we pass to the Fourier
transforms in the discrete but infinite-range indices and get
instead a continuous variable $z\in [-\half,\half]$. In the
notations of ref.~\cite{KvB}:
\beqa \label{Bmn}
\nonumber(B^\dag-x^\dag)_{nm}= -\int_{-1/2}^{1/2}\!\frac{dz}{2\pi  i}\,e^{-2\pi iz n}
\hat{D}^\dag_x(z)\,e^{2\pi izm},\qquad
\hat{D}^\dag_x(z)= -\d_z+\hat{A}^\dag(z)+2\pi
ix^\dag\equiv-\d_z+2\pi i r^\dag(z)\;,
\label{Dz}\eeqa
where
\beq \hat A(z)= 2\pi i[\xi+{\overrightarrow  \D}\cdot\vec\sigma\Theta_\omega(z)].
\label{eq:nahmdata}\eeq
Here the function
$\Theta_\omega(z)= 2\bar\omega$ when $z\in[-\omega,\omega]$ and
$-2\omega$ otherwise; $r^\dag(z)\equiv r_\mu(z)
\sigma^\dag_\mu,\;x^\dag\equiv x_\mu\sigma^\dag_\mu$. As it can be
seen from \eq{Dz}, the quaternion $\xi$ simply represents the
center of mass position of the whole system and can be set to
zero, $\xi= 0$. We define $r_\mu(z)= r_\mu$ when  $z\in [\omega,1-\omega]$, 
and $r_\mu(z)= s_\mu$ otherwise, where
\beq
\vec s= \vec x-2\bar\omega{\overrightarrow \D},\qquad
\vec r= \vec x+2\omega {\overrightarrow \D},\qquad s_4= r_4= x_4.
\label{vsvr}\eeq
Here $\bar\omega\equiv \half-\omega$, and $r_\mu$ and $s_\mu$ have the  meaning of the
vectors from the dyon centers to the `observation' point,
${\overrightarrow \D}= \vec r-\vec s$ has the meaning of dyon the separation. We choose
dyons to be separated in the 3d direction: ${\overrightarrow  \D}= \D\vec e_3$.
As for $\lambda$,
\beq
\lambda_n^\dag= \rho U^\dag e^{-2\pi i n\vec{\omega}\cdot\vec{\tau}} ,
\label{lambdan}\eeq
where $U$ is a unitary matrix, and $\rho > 0$.
We have an additional constraint~\cite{KvB}
\beq
{\overrightarrow \D}\cdot \vec\sigma
=\pi\rho^2(U^\dag\vec\omega\cdot\vec\sigma U)/\omega \;.
\eeq
Here $2\pi i \vec{\omega} \cdot  \vec{\tau} $ is the value of $A_4$ at
spatial infinity, $\omega= |\vec\omega|$. It can be seen that
$\pi\rho^2= \D $. We choose to rotate the $A_4$ direction in color
space instead of rotating monopole positions, so we do not loose
the generality of the solution. We connect the vector $\vec\omega$
and $U$ by
\beq
U^\dag \vec\omega\cdot\vec\tau U= \omega \tau_3 \;.
\eeq

Writing down the $m^{\rm th}$ component of the (infinite)
quaternionic vector as a Fourier transform
\beq \label{four}
u_m(x)= \int_{-1/2}^{1/2} u(x,z)e^{-2\pi i m z}dz,
\eeq
\eq{NahmEq} we have to solve can be rewritten as
\beq
(\d_z-2\pi i r^\dag(z))u(x,z)= 2\pi i\rho U^\dag\sum_n
e^{-2\pi i n \vec{\omega}\cdot\vec{\tau}}e^{2\pi i z n}
= 2\pi i\rho U^\dag (P_+\delta(z-\omega)+P_-\delta(z+\omega)),
\la{eqn}\eeq
where
\beq
P_\pm= \half(1\pm\vec\omega\cdot\vec \tau/\omega) \;.
\eeq
\Eq{eqn} is piece-wise homogeneous, therefore we present its
solution in the form
\beq
\label{u} u(x,z)\!= \!\left\{ \bea{ll} \exp\!\left(2\pi i s^\dag  z\right)
\!B_1,\quad &-\omega\!<\!z\!<\!\omega\\
\exp\!\left(2\pi i  r^\dag(z\!-\!1/2)\right)\!B_2,\;\;\;&\omega\!<\!z\!<\!1\!-\!\omega
\eea\right.
\eeq
and match the values and the derivatives of $u$ at the endpoints of the pieces,
\beq
e^{-2\pi i r^\dag\bar\omega} B_2-e^{2\pi i s^\dag\omega}  B_1= f_1,\qquad
e^{-2\pi i s^\dag \omega} B_1-e^{2\pi i r^\dag\bar\omega}B_2= f_2 \;,
\eeq
where $$f_1= 2\pi i\rho U^\dag P_+,\qquad f_2= 2\pi i\rho U^\dag  P_-,\qquad
\bar\omega= \half-\omega.$$ Note that $B_{1,2}$ are
matrices that generally do not commute:
\beqa\la{Bs}
&&B_2= \left(e^{-2\pi i s^\dag\omega}e^{-2\pi i r^\dag\bar\omega}
-e^{2\pi i s^\dag\omega}e^{2\pi i r^\dag\bar\omega}\right)^{-1}
\left(e^{-2\pi i s^\dag\omega}f_1 +e^{2\pi i
s^\dag\omega}f_2\right)\equiv b_{22} b_{21}e^{-2\pi i x_4
\omega\tau_3}U^\dag/\hat\psi,
\\
&&\nonumber B_1 = \left(e^{-2\pi i r^\dag\bar\omega}
e^{-2\pi i s^\dag\omega}
-e^{2\pi i r^\dag\bar\omega}e^{2\pi i s^\dag\omega}\right)^{-1}
\left(e^{-2\pi i r^\dag\bar\omega}f_2
+e^{2\pi i r^\dag\bar\omega}f_1\right)\equiv b_{12} b_{11}
e^{-2\pi i x_4 \omega\tau_3}U^\dag/\hat\psi ,
\eeqa
where
\beqa\nn
b_{22}&= &\left[-\cos(\pi x_4)(\ch\,\bsh\,\hat r+\bch\, \sh\,\hat s)
+i\sin(\pi x_4)(\ch\,\bch+\hat r\,\hat s\,\sh\,\bsh)\right],
\\
\nn
b_{12}&= &\left[-\cos(\pi x_4)(\ch\,\bsh\,\hat r+\bch\,\sh\,\hat s)
+i\sin(\pi x_4)(\ch\,\bch+\hat s\,\hat r\,\sh\,\bsh)\right],
\\
\nn b_{21}&= &2\pi i \rho (\ch-\hat s\,\tau_3\,\sh),\qquad
b_{11}= 2\pi i \rho (\bch+\hat r\,\tau_3\,\bsh)e^{\pi i x_4 \tau_3},
\\
\hat\psi&\equiv&-\cos(2\pi  x_4)+\ch\,\bch+\frac{\vec{s}\cdot\vec{r}}{s\,r}\sh\,\bsh \,.
\label{B12}\eeqa
Hat over the variable (notation found also in
\cite{KvB}) means contraction of the corresponding normalized
vector with Pauli matrices, e.g. $\hat{\omega}\equiv\vec\omega \cdot  \vec\tau/\omega $.
We denote for brevity
\beq
\shd\equiv\sinh(4\pi s\omega),\qquad \chd\equiv\cosh(4\pi  s\omega),\qquad
\bshd\equiv\sinh(4\pi r\bar\omega),\qquad \bchd\equiv\cosh(4\pi r\omega),
\eeq
and the hyperbolic functions with subscript ``$\half$''
are the corresponding functions of half the same arguments.
Combining eqs.(\ref{u},\ref{Bs},\ref{B12}) back into \eq{eq:vu} one
gets the two-dimensional quaternionic vector $v(x,z)$ which is the
base for the construction of the Green's function, see Appendix B.
Note that we have made a Fourier transform of $u$ \ur{four} and
got a continuous index $z$, so that scalar products of
infinite-dimensional vectors become $z$ integrations, see
\eq{brackets}.

We note that $U$ is actually a gauge transformation of $v$.
Therefore, the gauge potential $A_\mu^U$ is obtained by a global
gauge transformation of $A_\mu^{U= 1}$. We conclude that the
determinant does not depend on the relative `color orientation' of
the Polyakov line or holonomy, and of the vector $\overrightarrow{\D}$
connecting monopole centers. Thus, we set $U= 1$ and
$\vec{\omega}= \omega \vec e_3$.

We notice further that $v(x,z)$ built above gives $A_\mu$ that is
not periodic in time direction and zero $A_4$ at spatial infinity.
It is a peculiar feature of the `algebraic' gauge used in
\cite{KvB}. It is more convenient to use the gauge in which the
fields are periodic. To that end we make a non-periodic gauge
transformation $g= e^{2\pi i x_4 \omega \tau_3}$ and obtain
\beq
v(x,z)^{\rm per}= \Mphi^{-\half}(x)\left(\!\!\bea{c}-g\\w(x,z)
\eea\!\!\right),\;\;\;w\!= \!u g, \label{vbvper}\eeq meaning
\beq
w(x,z)= u(x,z)\;e^{2\pi i x_4 \omega\tau_3}.
\eeq

In terms of the Fourier-transformed $v$ the bracket takes the form
\beq
\langle v|\tilde v\rangle\equiv
v_1^\dag \tilde v_1+\int_{-1/2}^{1/2} v_2^\dag \tilde v_2\;dz \,,
\label{brackets}\eeq
where $v_1$ is an upper element and $v_2$ is a lower one.

Now let us determine $\Mphi(x)$. We use the following identities:
\beq
b_{12}^\dag b_{12}= b_{22}^\dag b_{22}= \hat\psi/2,\qquad
b_{21}^\dag  b_{21}= 4\pi^2\rho^2\left(\ch-\frac{s_3}{s}\,\sh\right),\qquad
b_{11}^\dag b_{11}= 4\pi^2\rho^2\left(\bch+\frac{r_3}{r}\,\bsh\right).
\la{4b}\eeq
Note that the right-hand sides of \eq{4b} are proportional to the unity
$2\times 2$ matrix. Now we can easily calculate the normalization:
\beqa \nn
\langle v| v\rangle &= & {\hat\psi}^{-2}b_{11}^\dag b_{11} b_{12}^\dag  b_{12}
\int_{-\omega}^{\omega}\!dz\,e^{-4\pi\vec{s}\cdot\vec\tau z}
+{\hat\psi}^{-2} b_{21}^\dag b_{21} b_{22}^\dag b_{22}
\int_{-\bar\omega}^{\bar\omega}\!dz\,e^{-4\pi\vec{r}\cdot\vec\tau z}\\
\nn
&= &\frac{\pi\rho^2}{\hat\psi}\left[\left(\frac{\bch\,\sh}{s}
+\frac{\ch\,\bsh}{r}\right)+\frac{\D}{s\,r}\sh\,\bsh\right]\equiv\frac{\psi-\hat\psi}{\hat\psi} \;.
\eeqa
We used the identity $\vec{r}-\vec{s}= \overrightarrow{\D}= \D\vec  e_3$. Thus for $\Mphi$
we get
\beq
\Mphi= \frac{\psi}{\hat\psi},\qquad
\psi= \hat\psi+\D\left(\frac{\bch\,\sh}{s}+\frac{\ch\,\bsh}{r}\right)
+\frac{\D^2}{s\,r}\sh\,\bsh\;.
\eeq

We have checked the $A_\mu$ of the KvBLL caloron \ur{APvB} by
calculating $\langle v^{\rm per}|\d_\mu v^{\rm per}\rangle$. Note
that $v^{\rm per}$ has the following periodicity property (only
for integer $n$):
\beq
v^{\rm per}(z,x_4+n)= e^{2\pi i n z}v^{\rm per}(z,x_4)\;.
\label{pln}\eeq

\section{Spin-0 isospin-1 propagator}

\subsection{General construction of the Green function}

Once the self-dual field is found in terms of the ADHM
construction, such that the gauge field is written as
$A_\mu= \langle v|\d_\mu v\rangle$ where the scalar product is
defined in \eq{brackets}, it is possible to construct explicitly
the Green function of spin-0 isospin-1 field in the background of
the self-dual field~\cite{Nahm80,KvB,LL}. The solution of the
equation
\beq
-\left(D_\mu^2\right)^{ca}(x)G^{ab}(x,y)= \delta^{cb}\,\delta^{(4)}(x-y)
\la{defGreen}\eeq
is given by
\beqa\label{green1}
G^{ab}(x,y)&= &\frac{\half\Tr\,\tau^a\langle v(x)|v(y)\rangle\tau^b
\langle v(y)|v(x)\rangle} {4\pi^2(x-y)^2}  \nonumber\\
&+&\frac{1}{4\pi^2}\int_{-1/2}^{1/2}dz_1\,dz_2\,dz_3\,dz_4\, M(z_1,z_2,z_3,z_4)
\half\Tr\!\left(\cv^{\dagger}(x,z_1) \cv(x,z_2)\tau^a\right)
\Tr\!\left(\cv^{\dagger}(y,z_4)\cv(y,z_3)\tau^b\right)\,,
\eeqa
where $\cv(x,z)\equiv \cB^\dag v(x,z)$.

We denote the first term by $G_1$ and the second term (the
M-part) by $G_2$. The only new object is the function
$M(z_1,z_2,z_3,z_4)$ which we determine bellow. As we shall see,
we do not need $M$ with arbitrary arguments, but only at
$z_3= z_4$. For coincident arguments we obtain
\beq
M(z_1,z_2,z,z)= \delta(z_1-z_2)M(z_1,z),
\eeq
see below.

The propagator (\ref{green1}) is written for the $\re^4$ space and
does not obey the periodicity condition. The periodic propagator,
however, can be easily obtained from it by a standard procedure:
\begin{equation}\label{greenPa}
{\cal G}(x,y)= \sum_{n= -\infty}^{+\infty}G(x_4,{\vec x};y_4+n,{\vec y}).
\end{equation}
In what follows it will be convenient to split it into three parts:
\beqa \nn
{\cal G}(x,y)= {\cal G}^\r(x,y)+{\cal G}^\s(x,y)+{\cal G}^\m(x,y),\\
\la{green3}
{\cal G}^\s\equiv \left.G_1\right|_{n= 0},\qquad{\cal G}^\r
\equiv\sum\limits_{n\neq 0}G_1,\qquad{\cal G}^\m\equiv\sum_n G_2 \;.
\eeqa
The vacuum current \ur{defJ} will be also split into
three parts, in accordance to which part of the periodic
propagator \ur{green3} is used to calculate it:
\beq
J_\mu= J^\r_\mu+J^\s_\mu+J^\m_\mu \;.
\eeq

\subsection{Propagator in the BPS dyon background}

In Appendix A.2 we have found the needed periodic quaternion
$v(x,z)$ for the single BPS monopole (see \eq{vmonostr}).
The 4-argument function $M$ for the BPS monopole was computed in
ref.~\cite{Nahm80}. The result with the two last arguments taken
equal is
\beqa
\nonumber &&M(z_3,z_4,z,z)= \delta(z_3-z_4)M(z_3,z),\\
&&M(z,z')= -\frac{1}{4\v^2}\left(2|z-z'|-1+4 z z'\right)
\label{Mmono} \;. \eeqa \Eqs{vmonostr}{Mmono} completely determine
the periodic propagator defined in \eqs{green1}{greenPa} in the BPS
dyon background. The use of this propagator is demonstrated in
Appendix C.

\subsection{Propagator in the KvBLL caloron background}

In Appendix A.3 we have found the needed quaternion
$v(x,z)$ for the KvBLL caloron. In this Appendix we derive the $M$-function
for the KvBLL caloron. The propagator \ur{green1} will be then
completely determined in the caloron background.

In the notations of \cite{Nahm80} $M$ is an infinite-dimensional
rank-4 tensor, with indices running from 1 to $k$, the topological
charge in $\re^4$. As in the case of $v$, it is convenient to make
the Fourier transformation with respect to the indices:
\beq
M_{pqnm}= \int_{-1/2}^{1/2}M(z_1,z_2,z_3,z_4)\,
e^{2\pi i (\!-\!p z_1\!+\!q z_2\!+\!n z_3\!-\!m z_4)}
dz_1dz_2dz_3dz_4.
\eeq
The tensor $M_{pqmn}$ is defined by the equation \cite{CWS}
\beq
\half\Tr[(\cA^\dag \cA)_{il}(\cB^\dag \cB)_{mj}+(\cB^\dag  \cB)_{il}(\cA^\dag \cA)_{mj}
-2(\cA^\dag \cB)_{il}(\cB^\dag  \cA)_{mj}]M_{rsij}=\delta_{rl}\delta_{sm}\;,
\label{nahmeq}\eeq
All indices here run from 1 to $k$ as rectangular $k\times (k+1)$ matrices $\cA$ and $\cB$
are contracted along the longer side. Here $\cA$ and $\cB$ are:
\beq
\Delta(x)\equiv \cA+\cB x,\quad\cA= \Delta(0),\quad
\cB=\left(\!\!\bea{c}0\\-1\eea\!\!\right)\;.
\label{ABvb}\eeq
\Eq{nahmeq} can be rewritten as
\beq
\label{92}
\half\Tr[(\Delta^\dag \Delta(0))_{il}\delta_{mj}+(\Delta^\dag  \Delta(0))_{mj}\delta_{il}
-2B^\dag_{il}B_{mj}]M_{rsij}= \delta_{rl}\delta_{sm} \;,
\eeq
where $\Delta^\dag \Delta(0)= \lambda^\dag\lambda+B^\dag B$, $\; B$
and $\lambda$ are found in \eq{Bmn} and \eq{lambdan},
respectively. In our case $k$ is infinite and we rewrite \eq{92}
in the Fourier basis:
\beq
\nonumber
\half\Tr\left[\tilde\Delta^\dag\tilde\Delta(0,z_3)
+\tilde\Delta^\dag\tilde\Delta(0,z_4)+ 2\left(\frac{\d_{z_3}}{2\pi i}
+r^\dag(z_3)\right)\left(\frac{\d_{z_4}}{2\pi  i}-r(z_4)\right)\right]M(z_1,z_2,z_3,z_4)
=\delta(z_1-z_3)\delta(z_2-z_4)\;,
\eeq
where $r(z)= r_i \sigma_i$ when $z\in [\omega,1-\omega]$, and $r(z)= s_i \sigma_i$
otherwise; $\sigma_i= i\tau_i$. Zero components of $r_\mu,\;s_\mu$
are absent because $x_\mu= 0$. We use
\beq
\tilde\Delta^\dag\tilde\Delta(0,z)= \!-\!\frac{\d_z^2}{4\pi^2}\!+\!r^2(z)
\!+\!\frac{\rho^2}{2}\left(\delta(z\!-\!\omega)\!+\!\delta(z\!+\!\omega)\right)\;.
\eeq
Here the first two terms come from the Fourier transformation
of $B^\dag B$ (\ref{Bmn}) and the last one comes from the Fourier
transformation of $\lambda^\dag\lambda$. We obtain the explicit
equation for $M$:
\begin{widetext}
\beqa &&\left(-\frac{(\d_{z_3}+\d_{z_4})^2}{4\pi^2}+|\vec{r}(z_3)
-\vec{r}(z_4)|^2\right)M(z_1,z_2,z_3,z_4)+\nonumber
\\
&& +\frac{\rho^2}{2}\left(\delta(\!z_3\!-\!\omega\!)
+\delta(\!z_3\!+\!\omega\!)+\delta(\!z_4\!-\!\omega\!)
+\delta(\!z_4\!+\!\omega\!)\right)  M(z_1,z_2,z_3,z_4)= \delta(z_1-z_3)\delta(z_2-z_4) \;.
\eeqa
\end{widetext}
In the case $z_3=z_4$, which is the only one we need as we shall
see in a moment, we look for the solution in the form
\beq
M(z_1,z_2,z,z)=\delta(z_1-z_2)M(z_1,z) \;.
\eeq
The equation for the two-argument function simplifies to
\beq
\left(-\frac{\d_{z}^2}{4\pi^2}+\frac{\D}{\pi}\delta(z-\omega)
+\frac{\D}{\pi}\delta(z+\omega)\right)M(z',z)= \delta(z-z') \;,
\eeq
where $\D=\pi \rho^2 $. We see that the solution has to be
piece-wise linear in its arguments. The solution is symmetric in
its two arguments and for $z<z'$ is given by
\begin{widetext}
\beqa \label{M} M(z,z')= \left\{\bea{ll}
\frac{32\D^2{\pi^3}\omega (z-\omega )(1-z'-\omega)
-8\D\pi^2 ({{\omega }^2}+z (z'-1))+\pi}
{2 d(8 \D \pi \omega \bar\omega+1)},\quad z,z' \in [\omega,1-\omega] \\
\frac{1}{2}\pi\left(\frac{4\pi z(1-2z')}{8\D\pi\omega\bar\omega+1}+
\frac{1}{\D}\right),\quad\quad\quad\quad\quad\quad\quad\quad\quad z \in [-\omega,\omega],
z'\in [\omega,1-\omega] \\
\frac{4\D\pi^2[z-z'-2z'z-2(\omega-1)\omega-8\D\pi(z'-\omega)
(z+\omega)\bar\omega]+\pi}
{2\D(8\D\pi\omega\bar\omega+1)},\quad z,z' \in [-\omega,\omega]
\eea \right.\;.
\label{Mvb} \eeqa
\end{widetext}
Outside this range $M$ is defined by periodicity:
$M(z+n,z'+m)= M(z,z')$, where $n,m$ are integers.

Now let us demonstrate that actually only the two-argument function
$M(z,z')$ is needed to construct the propagator satisfying the
periodicity. It turns out that making the Green function
periodic simplifies $G_2$ (section III). One has from the
definitions \ur{green1}-\ur{green3}:
\beq
{\cal G}^\m\equiv\sum_n
\frac{1}{8\pi^2}\int_{-1/2}^{1/2}\!dz_1\!\dots\!dz_4\,
M(\!z_1\!\dots\!z_4\!)\,\Tr\!\left(\!\cv^{\dagger}(x,z_1)
\cv(x,z_2)\tau^a\!\right)\Tr\!\left(\!\cv^{\dagger}(y^n,z_4)\cv(y^n,z_3)\tau^b\!\right) \;,
\eeq
where $y^n_4= y_4+n,\;\vec y^n= \vec y$. Using \eq{pln} we put
\beq
\cv(y^n,z)= e^{2\pi i\eta n z}\cv(y,z)
\eeq
Further on, we note that for $|\eta|\leq 1$ one has
\beq\nn
\sum_n\Tr\!\left(\cv^{\dagger}(y^n,z_4)\cv(y^n,z_3)\tau^b\right)
=\Tr\!\left(\cv^{\dagger}(y,z_4)\cv(y,z_3)\tau^b\right)\delta(z_3-z_4)\frac{1}{|\eta|}\;.
\eeq
Now we can see that making the Green's function periodic results in the substitution
\beq
\nonumber M(z_1,z_2,z_3,z_4)\rightarrow  \frac{1}{|\eta|}M(z_1,z_2,z_3,z_3)\delta(z_3-z_4)
= \frac{1}{|\eta|}M(z_1,z_3)\delta(z_1-z_2)\delta(z_3-z_4) \;.
\eeq
It follows from \eq{BPSpln} and \eq{pln} that for the
monopole one has to take $\eta=\v/(2\pi)<1$ and for the KvBLL
caloron $\eta=1$. In both cases the $M$-part of the periodic
propagator is given by
\beq
{\cal G}^\m=\frac{1}{8\pi^2|\eta|}\int_{-1/2}^{1/2}dz dz'  M(z,z')\,
\Tr\left(\cv^{\dagger}(x,z) \cv(x,z)\tau^a\right)\,
\Tr\left(\cv^{\dagger}(y,z')\cv(y,z')\tau^b\right),
\label{MtermFinal}\eeq
where the two-argument $M$ functions are found in \eq{Mmono} and \eq{Mvb},
respectively.

\section{Vacuum current in the BPS monopole background}

We compute the vacuum current \ur{defJ} in the BPS monopole
background in this Appendix. We assume $0<\v<2\pi$ and work in the
stringy gauge \ur{Mdyon} dropping the index $s$ in $v^s$ given by \eq{vmonostr}.

\subsection{Singular part of the monopole current $J^\s_\mu$}

This part of the current corresponds to the second term ${\cal
G}^\s$ in \eq{green3}. At $x\rightarrow y$ this part of the
propagator is singular. The regularization is presented in
Appendix E. \Eqs{Jsf}{ABmono} state:
\beq \label{mainj1}
{J^\s_{\mu}}^{ab}= i\varepsilon^{adb}
\tr\left(\tau^d j_{\mu}\right),\qquad
j_{\mu}= \frac{\v^2}{12\pi^2} \left\langle v\left| f
\sigma_{\mu}\Delta^{\dagger} f \right|v\right\rangle-{\rm h.c.},\qquad
\Delta^\dag(x)= i\partial_z+\v x^\dag\;.
\eeq
The function $f(z,z',x)$ for the BPS monopole is known~\cite{Nahm80}:
\begin{widetext}
\begin{eqnarray}\label{f}
f(x;z,z')=  -\frac{e^{i\v x_4(z-z')}}{2\v s}
\left(\sinh\v s |z-z'|+\coth\frac{\v s}{2}\,\sinh\v s z\,\sinh\v s z'
-\tanh\frac{\v s}{2}\,\cosh\v s z\,\cosh\v s z'\right).
\end{eqnarray}
\end{widetext}
Here we denoted by $s$ the distance to the M-monopole center.
It is  helpful to calculate the action of the Green function on
$v$. Since monopole is a static configuration, we can take
$x_4= 0$, moreover $f$ is a scalar function and we can move
$S^\dag_{-}$ matrix to the left:
\beq
\nonumber
\left.|\nu\>\equiv S_- f|v\>\right|_{x_4= 0}
= \frac{ \cosh(s v z)\tanh(s v/2)-2 z \sinh(s v z)}{4\sqrt{s v \sinh(s  v)}}1_2
+\frac{\sinh(s v z)\coth(s v/2)-2 z \cosh(s v z)}{4\sqrt{s v \sinh(s  v)}}\tau_3 \;.
\eeq
We use the following identities
\beqa
\nonumber&&S_{-}\vec n_r\vec\tau S_{-}^\dag= \tau_3 \;,\\
\nonumber&&S_{-}\vec n_\theta\vec\tau S_{-}^\dag= -\cos(\phi)\tau_1-\sin(\phi)\tau_2 \;,\\
&&S_{-}\vec n_\phi\vec\tau
S_{-}^\dag= \sin(\phi)\tau_1-\cos(\phi)\tau_2
\eeqa
and arrive, after simple algebra, to
\beq\nonumber
\{j_4,j_r,j_\theta,j_\phi\}=\frac{\v^3}{12 \pi^2}
\<\nu|\{i,-\tau_3,\cos\!\phi\;\tau_1\!+\!\sin\!\phi\;\tau_2,
-\!\sin\!\phi\;\tau_1\!+\!\cos\!\phi\;\tau_2\}(\d_z-\v\tau_3  s)|\nu\>+{\rm h.c.} \;.
\eeq
Finally we obtain the singular part of the vacuum current:
\beqa
\nonumber J^\s_r&= &0,\\
\nonumber J^\s_\phi&= & -\frac{i\v\left(s^2\v^2{\mathrm{csch}^2(s\v )
+ s\v\coth (s\v)-2}\right)}{24\pi^2 s^2\sinh(s\v)}(T_1\cos(\phi)+T_2\sin(\phi)),\\
\nonumber J^\s_\theta&= & -\frac{i\v\left(s^2\v^2{\mathrm{csch}^2(s\v )
+ s\v\coth (s\v)-2}\right)}{24\pi^2 s^2\sinh(s\v)}(T_1\sin(\phi)-T_2\cos(\phi)),\\
J^\s_4&= &-\frac{i\left( 1-s^3\v^3\coth(s\v){\mathrm{csch}^2(s\v)}\right)}{24\pi^2 s^3}T_3 \;,
\la{Jsmon}\eeqa
where
${(T_c)}^{ab}\equiv i\varepsilon^{acb}$.

\subsection{Regular part of the monopole current $J^\r_\mu$}

We are going to calculate the part of the current that corresponds to
\beq
({\cal G}^{\r})^{ab}(x,y)\equiv\sum_{n\neq 0}
\frac{1}{8\pi^2(x-y_n)^2}\Tr\left[\tau^a \<v(x)|v(y_n)\>\tau^b\<v(y_n)|v(x)\>\right],
\qquad y_n=\vec y,\quad y_{n4}=y_4+n,
\label{Gr}\eeq
namely
\beq
\nonumber
J^\r_\mu = J^{\r1}_\mu+J^{\r2}_\mu,\qquad
J^{\r1}_\mu = A_\mu {\cal G}^{\r} + {\cal G}^{\r}A_\mu,\qquad
J^{\r2}_\mu = (\d^x_\mu-\d^y_\mu){\cal G}^{\r} \;.
\eeq

At first consider $J^{\r1}_\mu$. We have to compute ${\cal
G}^{\r}$ with equal arguments. Substituting (\ref{vmonostr}) into
(\ref{Gr}) and calculating the trace one has
\beq
({\cal G}^{\r})^{ab}(x,x)\equiv\sum_{n\neq 0}\int dz  dz'\frac{\Tr}{8\pi^2n^2},
\label{Grxx}\eeq
where
\beq\nonumber
\Tr= 2\frac{s^2\v^2e^{i n\v(z-z')}}{\sinh^2(s\v)}
\left[\cosh(2s\v(z-z'))(\delta^{ab}-\delta^{a3}\delta^{b3})
+\cosh(2s\v(z\!+\!z'))\delta^a_3\delta^b_3\!+\!\sinh(2s\v(z\!-\!z'))T_3^{ab}\right]
\eeq
with $T^c= i\varepsilon^{acb}$. To compute the sum in this expression
we use the summation formula (note that $\v<2\pi$)
\beq
\sum_{n\neq 0}\frac{e^{i z n}}{4\pi^2
n^2}=  \frac{z^2}{8\pi^2}-\frac{
|z|}{4\pi}+\frac{1}{12},\qquad -2\pi<z<2\pi \;.
\eeq
It remains now to calculate integrals over $z$ and $z'$. The result is
\beqa
\nonumber
{{\cal G}^\r}(x,x)&= &\left[\frac{3 \coth(s\v )-s\v
(3\mathrm{csch}^2(s\v )+2)}{8 \pi s }
+\frac{(s \v \coth(s \v  )-1)^2}{8\pi^2s^2}\right](\delta^{ab}-\delta^a_3\delta^b_3)
\\
\nonumber &+&\left[\frac{s \v \;\mathrm{csch}^2(s\v)-\coth(s\v)}{8
\pi s }+\frac{1-s^2\v  ^2\;\mathrm{csch}^2(s\v)}{16\pi^2s^2}+\frac{1}{12}\right]\delta^{ab}\;.
\eeqa

Now we turn to the $J^{\r2}_\mu$ part of the current where
we have to sum over $n$ a derivative of the propagator. First of
all we consider derivatives of the trace in (\ref{Gr}). One finds
for $x= y$:
\begin{widetext}
\beqa \nn(\d^x_\theta-\d^y_\theta) \Tr &\!\!\!= &\!\!\!
\frac{2s^2\v^2 i}{\sinh(s\v)^2}(\cosh(2s\v z)+\cosh(2s\v z'))
(T_1\sin\phi- T_2\cos\phi)e^{i \v n(z-z')}\,,
\\
(\nonumber\d^x_\phi-\d^y_\phi) \Tr &\!\!\!= &\!\!\!
\left(\sin\theta\frac{4s^2\v^2i}{\sinh(s\v)^2}\cosh(s\v(z+z'))\cosh(s\v(z-z'))
( T_1\cos\phi+T_2\sin\phi)\right.
\\
\nn&&\!\!\!\left.-\frac{4s^2\v^2 i}{\sinh(s\v)^2}\cosh(2s\v(z-z'))
(1-\cos\theta)T_3\right)e^{i \v n(z-z')}\,,
\\
\nn(\d^x_4-\d^y_4)\Tr
&\!\!\!= &\!\!\!-\frac{4s^2\v^3i}{\sinh^2(s\v)}(z-z')\sinh(2s\v(z-z'))T_3 e^{i \v n(z-z')}\,,
\\
(\d^x_s-\d^y_s)\Tr &\!\!\!= &\!\!\!0 \;.
\eeqa
\end{widetext}
Here only terms even in $z-z'$ were left. The last two equations
are especially clear as we can drop out the matrices $S$ in
\eq{vmonostr}.

A derivative of the denominator of (\ref{Gr}) is equal to zero for
$x= y$ except for the derivative with respect to $x_4$, but in this
case we have the expression of the form (\ref{Grxx}) with
$n^3/4$ instead of $n^2$ in the denominator. Now we can sum over
$n$. We use the summation formula
\beq
\sum_{n\neq 0}\frac{e^{izn}}{i\pi^2 n^3}=\frac{z^3}{6\pi^2}
-\frac{z|z|}{2\pi}+\frac{z}{3},\;\;\;\;\;-2\pi<z<2\pi  \;.
\eeq
Next one has to integrate over $z,\;z'$. Combining all pieces we obtain:
\beqa \nn
J^\r_r&= &0,
\\
\nn  J^\r_\phi&= &i\frac{\cos(\phi)T_1+\sin(\phi)T_2}{48\sinh^3(s\v)\pi^2  s^2 }\wp_1,
\\
\nn
J^\r_\theta&= &i\frac{\sin(\phi)T_1-\cos(\phi)T_2}{48\sinh^3(s\v)\pi^2  s^2 }\wp_1,
\\
J^\r_4&= &\frac{iT_3}{24{\pi }^2s^3}\wp_2,
\la{Jrmon}\eeqa
where we denote
\beqa
\nn \wp_1&\equiv&(s^2\v^3+6\pi s^2\v ^2+3\v+3s(\v+\pi)\sinh(2s\v )\v
-(s^2\v ^3 +3\v  +6\pi)\cosh(2s\v )+6\pi),
\\
\nn\wp_2&\equiv& 8{\pi }^2s^2(-1+s\v \coth(s\v))
- 12\pi s\coth(s\v){(-1+\Mfunction{s}\v\coth(s\v))}^2\\
\nn
&+&(-3(1+4s^2\v^2)+s\v(4(3+s^2\v^2)\coth(s\v)
+3s\v(-4+s\v\coth(s\v)){\mathrm{csch}^2(s\v)})).
\eeqa
We have used spherical coordinates. For example, a projection of
$\vec J$ onto the direction
$\vec{n}_\theta= (\cos\theta\cos\phi,\cos\theta\sin\phi,-\sin\theta)$
is denoted by $J_\theta$.

\subsection{M-part of the monopole current $J^\m_\mu$ }

Combining together \eqs{Mmono}{MtermFinal} we have for the
$M$-part of the periodic Green's function:
\beq \label{MmonoAp}
{\cal G}^\m=-\frac{\v}{16\pi}\int_{-1/2}^{1/2}\!dz dz' \!
\left(2|z\!-\!z'|\!-\!1\!+\!4 z z'\right)\,
\Tr\!\left(\!v^{\dagger}(x,z) v(x,z)\tau^a\!\right)\,
\Tr\!\left(\!v^{\dagger}(y,z')v(y,z')\tau^b\!\right)
\eeq
Note that we can drop out $S_-$ in (\ref{vmonostr}). In the stringy
gauge one has
\beq
{v}^\dag(x_\mu,z)v(x_\mu,z)=\frac{\v s}{\sinh(\v s)}\exp[-2\v s\tau^3 z] \;.
\eeq
It means that ${\cal G}^\m$ has only the `33' component. Taking the trace we get:
\beq
\Tr\left[{v}^\dag(x_\mu,z)v(x_\mu,z)\tau^3\right]=-2\frac{\v s}{\sinh(\v s)}\sinh(2\v s z) \;.
\eeq
 Therefore the only nonzero component of ${\cal G}^\m$ is
\beq
\nn
{\cal G}^{33}_\m(x,y)= -\frac{\v^3 r_x r_y}{4\pi}\int_{-1/2}^{1/2}\! dz dz'
\left(2|z\!-\!z'|\!-\!1\!+\!4 z z'\right)\,
\frac{\sinh(2\v r_y z')\sinh(2\v r_x z)}{\sinh(\v r_x)\sinh(\v r_y)} \;,
\eeq
($r_x= |\vec{x}|$, $r_y= |\vec{y}|$).
Performing the integrations we get
\beq
{\cal G}^{33}_\m(x,y)= -\frac{1}{4\pi\v }\left(\frac{1}{r_x r_y}
+\frac{\v r_x \coth \v r_y-\v r_y\coth\v r_x}{r^2_y-r^2_x}\right) \,.
\label{Gmmono}\eeq
Note that (\ref{Gmmono}) is symmetric in its arguments. For that reason  the contribution to
the current coming from (ordinary) derivatives of ${\cal G}^\m$ is zero,
$$\left . \partial_x {\cal G}^\m(x,y)-\partial_y{\cal G}^\m(x,y)\right|_{x=y}=0, $$
and only the anticommutator $\{{\cal G}^\m(x,x),A_\mu\}$
remains. Taking the limit $x \rightarrow y$ we get for ${\cal G}^\m$
\beq
{\cal G}^{33}_\m(x,x)= \frac{1}{8\pi  s}
\left[\coth(\v s)-\frac{2}{\v s}+\frac{\v s}{\sinh^2(\v s)}\right]
\eeq
and for the contribution to the current
$J^\m_\mu=A_\mu{\cal G}^\m+{\cal G}^\m A_\mu$ we obtain in spherical coordinates
\beqa
\nn&J^\m_r=0,&\\
\nn&J^\m_\phi=-\frac{i(\sin(\phi)T_2+\cos(\phi)T_1)\left(s\v \coth(s\v) +
{\Mfunction{s}}^2\v ^2{\mathrm{csch}^2(s\v)}-2\right) }{16\pi \sinh(s\v)s^2},&\\
\nn&J^\m_\theta=-\frac{i(\sin(\phi)T_1-\cos(\phi)T_2)\left(s\v \coth(s\v) +
{\Mfunction{s}}^2\v ^2{\mathrm{csch}^2(s\v)}-2\right)}{16\pi \sinh(s\v)s^2},&\\
&J^\m_4=0 \;. &
\la{Jmmon}\eeqa

Adding up \urs{Jsmon}{Jrmon} and \ur{Jmmon} we obtain the full vacuum  current in the
BPS background, see \eq{monopoleCurrent} of the main text.

\section{Vacuum current in the KvBLL caloron background}

There are no principal problems to make the calculation of the
caloron Green's function and the ensuing vacuum currents exactly.
One can consider this Appendix as an instruction how to perform
the exact calculation. In fact, we have done it but unfortunately
the exact result for the current is about 200 pages long and thus
too large to be printed.  However, in certain limits the
expressions drastically simplify. In particular, assuming the case
when the dyons inside the caloron are widely separated such that
their cores do not overlap, it is relatively easy to find the
KvBLL caloron current with the exponential precision (i.e.
dropping out term of the order $e^{-r\bv},\;e^{-s\v}$). This will
be sufficient to find the determinant of the KvBLL caloron for
large $\D$ up to some constant.

With the exponential precision, the only nonzero components of the KvBLL  caloron's gauge
potential in fundamental representation are (see section II)
\beq
A_4\simeq\frac{i\tau_3}{2}\left(4\pi\omega+\frac{1}{r}-\frac{1}{s}\right),\qquad
A_\varphi\simeq-\frac{i\tau_3}{2}\left(\frac{1}{r}+\frac{1}{s}\right)
\sqrt{\frac{(\D-r+s)(\D+r-s)}{(\D+r+s)(r+s-\D)}}\;.
\label{EAvb}\eeq
We are using the coordinates $x_4,r,s,\varphi$, where $r,s$ are defined
in (\ref{vsvr}) and $\varphi$ is defined by
\beq
\vec x = x_\varrho(\cos\!\varphi\;\vec e_2+\sin\!\varphi\;\vec e_1)
\!+\!\left(\!\frac{\D^2+r^2-s^2}{2\D}-2\D\omega\right)\vec e_3,\quad
x_\varrho\equiv\frac{\sqrt{(\D\!+\!r\!-\!s)(\D\!+\!s\!-\!r)(r\!+\!s\!-\!\D)
(\D\!+\!r\!+\!s)}}{2\D}.
\eeq
One can easily check the consistency of this definition, i.e. that
\beq
\nn s=|\vec s|,\;\;r=|\vec r|,\;\;{\rm where}\;\;
\vec s=\vec x-2\bar\omega{\overrightarrow \D},\;\;\;\vec r=\vec x+2\omega
{\overrightarrow \D},\;\;\;{\overrightarrow \D}=\D\vec e_3 \;.
\eeq
Since $A_r=0$ and $A_s=0$ we have to calculate only the $J_4$ and $J_\phi$
components.

We shall use the ADHM construction. The main steps of the
calculation are similar to that for the monopole. Dropping out
exponentially small terms in \eq{vbvper} one has in the
periodic gauge
\beq
v(x,z)\simeq\sqrt{\frac{r+s-\D}{r+s+\D}}
\left(\!\!\bea{c}
-e^{2\pi i x_4\omega\tau_3}\\w(x,z)
\eea\!\!\right),
\la{Evbvper}\eeq
\begin{widetext}
\beq w(x,z)\simeq\left\{\bea{cc}
-i\pi\rho e^{2\pi (i s^\dag z-\omega s)}\left[\frac{\D+r-s}{\D}\tau_3
+\frac{2 x_\varrho}{r+s-\D}
\left(\tau_1\sin\!\varphi+\tau_2\cos\!\varphi\right)\right],\;\;\;\;\;&\!
-\!\omega\!<\!z\!<\!\omega\\
{\bea{c} i\pi\rho e^{2\pi (i r^\dag z-i r^\dag/2-r\bar\omega)}
\left[\frac{\D+s-r}{\D}\left(i1_2\sin\!\pi x_4 +\tau_3\cos\!\pi x_4\right)+\right.\\
\left.+\frac{2 x_\varrho}{r+s-\D}\left(\sin(\pi x_4\!-\!\varphi)\tau_1\!
-\!\cos(\pi x_4\!-\!\varphi)\tau_2\right)\right]
\eea},\;\;\;\;\;&\!\omega\!<\!z\!<\!1\!-\!\omega \; \eea \right.
\eeq
\end{widetext}
where $s^\dag\equiv s_\mu\sigma^\dag_\mu,\;s\equiv |\vec s|,\;s_4= x_4$.
We shall use the following formulas to pass to the
cylindrical coordinates $x_3,x_\varrho,\varphi$:
\beq
\frac{\d r}{\d x_3}= \frac{\D^2+r^2-s^2}{2\D r},
\;\;\;\;\;\frac{\d r}{\d  x_\varrho}= \frac{x_\varrho}{r},\;\;\;\;\;\frac{\d r}{\d \varphi}=0\qquad
\frac{\d s}{\d x_3}= -\frac{\D^2+s^2-r^2}{2\D s},
\;\;\;\;\;\frac{\d s}{\d x_\varrho}= \frac{x_\varrho}{s},\;\;\;\;\;\frac{\d s}{\d \varphi}=0\;.
\eeq

\subsection{Singular part of the caloron current $J^\s_\mu$}

Let us calculate the singular part of the vacuum current with
exponential precision. It is related to the zero Matsubara
frequency. Similar to the monopole case, we could use \eq{Jsf},
where the Green's function (\ref{deff}) for the case of KvBLL
caloron was found in~\cite{KvB}. However it is more convenient to
use \eq{jour}) because then we have only to take derivatives of
the simple expression \ur{Evbvper} and no integrations arise.
\Eq{Jsf}  would have been more suitable for the exact calculation.

It is straightforward to calculate the quantity $\Gamma$ from
\eq{Gamma3}. It is sufficient to calculate the second time
derivative:
\beq
\Gamma\simeq<\d_4\d_4 v|v>-A_4^2 \;.
\eeq
Bearing in mind that $\Gamma$ is a vector under gauge transformations, we
can perform calculations in any gauge. Up to the exponentially
small terms we have
\beq
\Gamma^{ab}= -\frac{\left(r+s\right)
\left({\left(r-s\right)}^2+\D\left(r+s\right)\right)
\delta^{ab}}{4r^2s^2\left(\D+r+s\right) } \;.
\eeq
One can observe from \eq{EAvb} that all terms with
derivatives in the right-hand side of \eq{jour} are zero. Writing
the Laplace operator in the cylindrical coordinates we find
\beqa
\nn
j^\s_4\!&\simeq&\!\frac{1}{48\pi^2}\!\left[\<\d_4\!\left(\d_4^2\!+\!\d_3^2\!
+\!\frac{1}{x_\varrho}\d_\varrho x_\varrho \d_\varrho\!
+\!\frac{1}{x_\varrho^2}\d_\varphi^2\right)\! v|v\>\!-\!{\rm h.c.}\right]
+\frac{1}{24\pi^2}\!\left(A_4^3\!+ \!A_\varphi A_4 A_\varphi\!+\!6A_4\Gamma\right)\,,
\\
\nn
j^\s_\varphi\!&\simeq&\!\frac{1}{48\pi^2}\!\left[\<\frac{\d_\varphi}{x_\varphi}\!
\left(\d_4^2\!+\!\d_3^2\!+\!\frac{1}{x_\varrho}\d_\varrho x_\varrho \d_\varrho\!
+\!\frac{1}{x_\varrho^2}\d_\varphi^2\right)\! v|v\>\!-\!{\rm h.c.}\right]
+\frac{1}{24\pi^2}\!\left(A_\varphi^3\!+\!A_4 A_\varphi A_4 \!+\!6A_\varphi\Gamma\right)\;.
\eeqa
Taking the derivatives we obtain simple expressions:
\beqa
\la{j4cals}
j^\s_4 &=& \frac{i\tau_3}{48\pi^2}\left(\frac{1}{r^3}-\frac{1}{s^3}\right),\\
\la{jphicals}
j^\s_\phi &=&
-\left(\frac{1}{r}+\frac{1}{s}\right)\frac{i\tau_3x_\varrho \D}{8\pi^2 r s (\D+r+s)^2} \;.
\eeqa

\subsection{Regular part of the caloron current $J^\r_\mu$}

Next we calculate the temperature-dependent part of the KvBLL
caloron vacuum current. As in the monopole case (Appendix C.2)
we divide the current into two parts,
\beq
J^\r_\mu= J_\mu^{\r2}+J_\mu^{\r1} \;,
\label{dJr0}\eeq
where
\beqa
\nn 
J_\mu^{\r2}&= &\sum_{n\neq 0}\frac{1}{8\pi^2}(\d_\mu^x-\d_\mu^y)
\frac{\Tr[\tau^{a} v^\dag(x)v(y)\tau^{b} v^\dag(y)v(x)]}{(x-y)^2},\\
\nn 
J_\mu^{\r1}&= &\sum_{n\neq 0}\frac{1}{8\pi^2}
\left\{A_\mu,\frac{\Tr[\tau^a v^\dag(x)v(y)\tau^b
v^\dag(y)v(x)]}{n^2}\right\}
\eeqa
and $y_4= x_4+n$. The quaternion function $v(x,z)$ has been constructed
in Appendix A.3 (actually called $v^{\rm per}(x,z)$ there). It is important that
$v(x,z)$ has the remarkable periodicity property \ur{pln}.

In evaluating the above currents the tactics is to factor the matrix  part out of
the integrals over $z$. We use the following notations for the integrals over $z$:
\beqa
\nn I^n_+ &\equiv & \int_{-\omega}^\omega e^{2\pi i n z}\cosh(4\pi s z)dz,
\qquad\bar I^n_+\equiv  \int_{-\bar\omega}^{\bar\omega} e^{2\pi i n (z+1/2)}\cosh(4\pi r z)dz,\\
\nn I^n_- &\equiv & \int_{-\omega}^\omega e^{2\pi i n z}\sinh(4\pi s z)dz,
\qquad\bar I^n_- \equiv \int_{-\bar\omega}^{\bar\omega} e^{2\pi i n (z+1/2)}\sinh(4\pi r z)dz \;.
\eeqa
We obtain the following relations for the matrix structures:
\beqa
&&\nn\hat\psi^2 w^\dag(x_4) w(x_4+n)= (I^n_+ \beta+I_-^n \beta_s)
+(\bar  I^n_+\bar\beta+\bar I_-^n\bar\beta_r),\\
&&\nn\hat\psi^2 w^\dag(x_4+n) w(x_4)= (I^n_+\beta-I_-^n \beta_s)
+(\bar  I^n_+\bar\beta-\bar I_-^n\bar\beta_r),\\
&&\nn\hat\psi^2 w^\dag(x_4)\bar\d_4 w(x_4+n)
=(I^n_+\beta_0+I_-^n \beta^0_s+i\d_s I_-^n \beta+i\d_s I_+^n \beta_s)+...\\
&&\nn\hat\psi^2 w^\dag(x_4+n)\bar\d_4 w(x_4)
=(I^n_+ \beta_0-I_-^n \beta^0_s-i\d_s I_-^n \beta+i\d_s I_+^n \beta_s)+...
\eeqa
where `...' means the same expression but with
bar over each quantity and $r$ instead of $s$. The notation
$\bar\d_4$ means: derivative from the right minus derivative from
the left. The definition and the evaluation of the matrix structures with the
exponential precision is
\beqa
\nn
\beta^0_s&\equiv &b_{11}^\dag b_{12}^\dag \hat s\overline{\d}_4 b_{12} b_{11}
\simeq  o(e^{4\pi s\omega} e^{8\pi r\bar\omega}),\\
\nn
\bar\beta^0_r&\equiv &b_{21}^\dag b_{22}^\dag \hat r\overline{\d}_4b_{22} b_{21}
\simeq  \frac{i\pi^2 \D }{2}(\vartheta+1)(s_3-1)e^{8\pi s\omega}e^{4\pi  r\bar\omega},\\
\nn
\beta_0&\equiv &b_{11}^\dag b_{12}^\dag\overline{\d}_4 b_{12} b_{11}
\simeq  o(e^{4\pi s\omega} e^{8\pi r\bar\omega}),\\
\nn
\bar\beta_0&\equiv &b_{21}^\dag b_{22}^\dag\overline{\d}_4 b_{22}b_{21}
\simeq \frac{\pi^2 \D}{2i}(\vartheta+1)(s_3-1)\hat{\omega}e^{8\pi s\omega}e^{4\pi r\bar\omega},\\
\nn
\beta_s&\equiv &b_{11}^\dag b_{12}^\dag\hat{s} b_{12} b_{11}
\simeq \frac{\pi \D}{4}(\vartheta+1)(r_3+1)\hat\omega e^{4\pi s\omega}e^{8\pi r\bar\omega},\\
\nn
\bar\beta_r&\equiv &b_{21}^\dag b_{22}^\dag\hat{r} b_{22} b_{21}
\simeq \frac{\pi \D}{4}(\vartheta+1)(s_3-1)\hat\omega e^{8\pi  s\omega}e^{4\pi r\bar\omega},\\
\nn
\beta&\equiv &b_{11}^\dag b_{12}^\dag b_{12} b_{11}
\simeq \frac{\D\pi}{4}(\vartheta+1)(r_3+1)e^{4\pi s\omega} e^{8\pi r\bar\omega},\\
\nn
\bar\beta&\equiv &b_{21}^\dag b_{22}^\dag b_{22} b_{21} \simeq
\frac{\D\pi}{4}(\vartheta+1)(1-s_3)e^{8\pi s\omega} e^{4\pi r\bar\omega},
\eeqa
where
\beq \nn
\hat{\psi} \simeq \frac{1}{4}(\vartheta+1)e^{4\pi s\omega}e^{4\pi r\bar\omega},
\qquad\vartheta\equiv\frac{\vec{r}\,\vec{s}}{s r}
=\frac{r^2\!+\!s^2\!-\!\D^2}{2s r},\qquad
r r_3=s s_3=\frac{\D^2\!+\!r^2\!-\!s^2}{2\D} \;.
\eeq
Substituting this into in the currents $J^{r_1,r_2}_\mu$ we obtain
certain sums, which are of the form
\beq\nn
\sum \limits_{n\neq 0}I_+^nI_+^n/(4\pi^2n^2),\qquad
\sum \limits_{n\neq 0}\cos(2\pi n\omega)g I_+^n/(4\pi^2n^2),\qquad
\sum\limits_{n\neq 0}\sin(4\pi n\omega)/(\pi^2n^3).
\eeq
All such sums can be calculated using the summation formulae
\beqa
\nn&&\sum_{n\neq 0}\frac{e^{2\pi i n (z-1/2)}}{4\pi^2 n^2}=\frac{z^2}{2}-\frac{1}{24}
\equiv c_2(z),\;\;\; -\half\!\leq\!z\!\leq\!\half\,,\\
\nn &&\sum_{n\neq 0}\frac{e^{2\pi i n (z-1/2)}}{\pi^2 n^3}
=8\pi i\left(\frac{z^3}{6}-\frac{z}{24}\right)
\equiv c_3(z),\;\;\; -\half\!\leq\! z\!\leq\! \half \;.
\eeqa
For example,
\beq \nn
\sum_{n\neq 0}I_+^nI_+^n/(4\pi^2n^2)
=\int_{-\omega}^{\omega}\int_{-\omega}^{\omega}c_2(z+z'-1/2)\cosh(4\pi s z)\cosh(4\pi s z')\;dz\;dz'
\eeq
and so on. With some help from {\tt Mathematica} we come to the final result
\begin{widetext}
\beqa
\label{J4calr}
J^\r_4=\left[\frac{1}{4{\pi }^2\,r^3}
-\frac{1}{{\pi }^2\,r^2\,s} + \frac{1}{{\pi }^2\,r\,s^2} -
\frac{1}{4\,{\pi }^2\,s^3} - \frac{1}{\pi \,r^2} + \frac{2}{\pi
\,r\,s} - \frac{1}{\pi \,s^2} + \frac{2}{3\,r} - \frac{2}{3\,s}+\right. \\
\nonumber
\left.\left(\frac{4}{\pi \,r^2}
- \frac{8}{\pi \,r\,s} + \frac{4}{\pi \,s^2} - \frac{8}{r}
+\frac{8}{s} + \frac{8\,\pi }{3}\right)\,\omega
+\left(\frac{16}{r}-\frac{16}{s}-16\,\pi\right)\,{\omega }^2
+\frac{64\,\pi\,\omega^3}{3}\right]\frac{i T_3}{2},
\eeqa
\beq
\label{Jphicalr}
J^\s_\phi=\left(\frac{1}{r}+\frac{1}{s}\right)\frac{i T_3x_\varrho \D}{4\pi^2 r s (\D+r+s)^2} \;.
\eeq
\end{widetext}

\subsection{M-part of the caloron current $J^\m_\mu$}

This part of the current is especially simple: with exponential precision it is zero.
The main steps are the same as in the case of a single monopole.
The starting formula is our \eq{MtermFinal}. First of all we note
that only the lower components of $v$ are left and only the $a=3$
component is nonzero:
\beq\nn
\Tr\left[\cv^+(x,z)\cv(x,z)\tau^a\right]
=\frac{1}{\phi(x)}\Tr\left[w^+(x,z)w(x,z)\tau^a\right]\propto\delta^{a3} \;.
\eeq
Inspecting the definition of the M-part of
the propagator \ur{MtermFinal} we observe that
\beq
G^{\m\,ab}(x,y)\propto \delta^{a3}\delta^{b3},\;\;\;\;\;
G^{\m\,ab}(x,y)= G^{\m\, ab}(y,x) \; .
\eeq
The second equation means that the terms with derivatives in the expression
for the current \ur{defJ} cancel each other. It follows from the first one that
the product of $G^\m $ and $A_\mu^{ab}\propto \varepsilon^{3ab}$
is equal to zero, too. Therefore we conclude that
\beq
J^\m_\mu\simeq 0 \;.
\eeq

\section{Regularization of the current}

Here we consider in more detail $J^\s_\mu$, the contribution to
the current from the singular (as $x \rightarrow y$) part of the
propagator ${\cal G}^\s(x,y)$ defined by \eq{green3}. This part
is obviously temperature-independent, so the zero-temperature
results are applicable. We regularize the current by setting
$x-y=\epsilon$ and inserting a parallel transporter to support
gauge invariance, see e.g.~\cite{BC}:
\beqa
\nn
J_\mu^\s\;&= &J_\mu^{\s1}+J_\mu^{\s2},\\
\nonumber
J_\mu^{\s1}&\equiv& \left[A_\mu(z-\epsilon/2)
G^\s(z-\epsilon/2,z+\epsilon/2)\right.
+\left.G^\s(z-\epsilon/2,z+\epsilon/2)A_\mu(z+\epsilon/2)\right]
\mathrm{Pexp}\left(-\int_x^y A_\mu dz_\mu\right),
\label{Jsmon1}
\\
J_\mu^{\s2}&\equiv&\left[(\partial^x_\mu-\partial^y_\mu)G^\s(x,y)\right]
\mathrm{Pexp}\left(-\int_x^y A_\mu dz_\mu\right) \;,
\label{Jsmono}\eeqa
where $x= z-\epsilon/2,\;y= z+\epsilon/2$ and we imply averaging over
all directions of $\epsilon$ in the $4d$ space. This regularization method
was proved to be equivalent to the $\zeta$-function regularization approach~\cite{CGOT}.

For a background field written in terms of the ADHM construction,
a useful expression for the vacuum current was derived in
refs.~\cite{CGOT,BC}. In the $SU(2)$ case it acquires the form:
\beq \nn
{J^\s_{\mu}}^{ab}=i\varepsilon^{adb}\tr\left(\tau^d j_{\mu}\right),\qquad
j_{\mu}=\frac{1}{12\pi^2} \left\langle v\left|\cB f
\left(\sigma_{\mu}\Delta^{\dagger}\cB-\cB^\dag\Delta\sigma_{\mu}^{\dagger}\right)
f \cB^\dag\right|v\right\rangle
\label{Jsf}\eeq
(see Appendix A for notations of the ADHM construction elements).

We would like to derive another expression for this part of the
current -- in terms of derivatives. In some cases it is more
useful. We start from writing our result:
\beq
j_\mu= \frac{1}{48\pi^2}\left[\left(D_\mu D^2\<v|\right)|v\> -
{\;\rm h.c.}\right]\;.
\label{ourj}\eeq
Let us prove it. First of all we consider the action of one derivative
\beqa
\nn
D_\mu\<v(x)|&\!\!\!= &\!\!\!\d_\mu\<v|-\d_\mu\<v|v\>\<v|= \d_\mu\<v|(1-|v\>\<v|)\\
&\!\!\!= &\!\!\!\d_\mu\<v|\Delta f\Delta^\dag = -\<v|\d_\mu\Delta f\Delta^\dag\\
&\!\!\!= &\!\!\!-\<v|\cB \sigma_\mu f\Delta^\dag \;.
\eeqa
At the end of the first line we have used \eq{vv2f}. The first equation
in the second line comes from differentiating the ADHM equation
$$0= \d_\mu(\<v|\Delta)= \d_\mu\<v| \Delta + \<v| \d_\mu \Delta.$$
The last equation follows from the definition (\ref{ABvb}).
Therefore we obtain
\beq\nn
D_\mu \<v(x)|v(y)\>= -\<v(x)|\cB \sigma_\mu  f_x\Delta_x^\dag|v(y)\>
=-\<v(x)|\cB \sigma_\mu f_x (x-y)^\dag\cB^\dag|v(y)\>\;,
\la{Dmuvv}\eeq
where in the last line we have used the ADHM equation
\ur{deltav}. We next consider two derivatives. It is
important here that $f$ is proportional to the unity $2\times 2$
matrix. We have
\beqa
\nonumber
D^2_x\<v(x)|v(y)\>&\!\!\!= &\!\!\!-D^x_\mu\left(\<v(x)|\cB \sigma_\mu
f_x (x-y)^\dag\cB^\dag|v(y)\>\right)\\
\nonumber &\!\!\!= &\!\!\!\<v(x)|\cB \sigma_\mu f_x\Delta_x^\dag\cB
\sigma_\mu f_x (x-y)^\dag\cB^\dag|v(y)\>-\<v(x)|\cB \sigma_\mu \d_\mu
f_x (x-y)^\dag\cB^\dag|v(y)\>-\<v(x)|\cB \sigma_\mu f_x  \sigma^\dag_\mu\cB^\dag|v(y)\>\\
\nonumber
&\!\!\!= &\!\!\!-4\<v(x)|\cB f_x\cB^\dag|v(y)\> \;.
\eeqa
We have used here
\beq
\nonumber
\sigma_\mu\d_\mu f_x=-f_x\d_\mu(\sigma_\mu\Delta^\dag\Delta)f_x
=-f_x\sigma_\mu(\sigma_\mu^\dag\cB^\dag\Delta+\Delta^\dag\cB\sigma_\mu)f_x
=-2f_x\cB^\dag\Delta f_x=f_x\sigma_\mu\Delta^\dag\cB\sigma_\mu f_x \;.
\eeq
We have also used that the derivative of the inverse operator is
$\d (O^{-1})=-O^{-1}(\d O)O^{-1}$, as well as the relations
\beq
\sigma_\mu\sigma^\dag_\mu=4,\;\;\;\;\;\sigma_\mu c\;\sigma_\mu=-2c^\dag,
\eeq
where $c$ is an arbitrary quaternion.

Finally, let us consider three derivatives:
\beq
\nonumber
D_\mu^xD^2_x \<v(x)|v(y)\>=-4D_\mu^x
\left(\<v(x)|\cB f_x\cB^\dag|v(y)\>\right)
=4\<v(x)|\cB f_x\sigma_\mu\Delta_x^\dag\cB
f_x\cB^\dag|v(y)\>+4\<v(x)|\cB \d_\mu f_x\cB^\dag|v(y)\>  \;.
\eeq
Notice that the last term is hermitian at $x=y$. Thus we have
proven that the current written in form of \eq{ourj} is equivalent
to that of \eq{Jsf}:
\beq \nn
\frac{1}{48\pi^2}\left[\left(D_\mu D^2\<v|\right)|v\> - {\;\rm h.c.}\right]
=\frac{1}{12\pi^2}\left\langle v\left| \cB f
\left(\sigma_\mu\Delta^\dag\cB-\cB^\dag\Delta\sigma^\dag_\mu\right)
f\cB^\dag\right|v\right\rangle \;.
\eeq

In fact it is more useful to rewrite everything in terms of
ordinary rather than covariant derivatives:
\beq
\left(D_\mu D^2\<v|\right)|v\>= \<\d_\mu\d^2 v|v\>+A_\nu A_\mu A_\nu
-\d_\nu A_\nu A_\mu- A_\nu \d_\nu A_\mu - \d_\mu A_\nu A_\nu
+\d_\mu \d_\nu A_\nu+ 6A_\mu\Gamma \;,
\la{GammaMu}\eeq
where $A_\mu$ is in the fundamental representation and
\beq
\half\left(D_\mu D_\nu+D_\nu D_\mu\<v|\right)|v\>
=\delta_{\mu\nu}\Gamma\;.
\la{dDwxy}\eeq
We have to prove that the left-hand-side of \eq{dDwxy} is a Lorentz
scalar as is the right-hand side. Note that \eq{Dmuvv} is
proportional to $x-y$. The only way to obtain a nonzero result at
$x-y\to 0$ is to differentiate this factor:
\beq
\half\left(D_\mu D_\nu+D_\nu D_\mu\<v|\right)|v\>
=-\half\<v|\cB(\sigma_\mu\sigma^\dag_\nu+\sigma_\nu\sigma^\dag_\mu) f \cB^\dag|v\>
=-\delta_{\mu\nu}\<v|\cB f \cB^\dag|v\>\;.
\la{Gamma1}\eeq
It follows from \eq{Gamma1} that $\Gamma$ is hermitian. We can
write $\Gamma$ as follows:
\beq
\Gamma\delta_{\mu\nu}= \<\d_\mu\d_\nu v|v\>+\frac{1}{2}(\d_\mu  A_\nu+\d_\nu A_\mu)
-\half(A_\mu A_\nu+A_\nu A_\mu)\;.
\label{Gamma3}\eeq
Finally, the regularized singular part of the current can be written as
\beq \label{jour}
j_\mu=\frac{1}{48\pi^2}\left(\<\d_\mu\d^2 v|v\>-\;{\rm  h.c.}\right)
+\frac{1}{24\pi^2}\left(A_\nu A_\mu A_\nu  + \d_\mu \d_\nu
A_\nu + 3A_\mu\Gamma+ 3\Gamma A_\mu\right) \;.
\eeq

Eqs.(\ref{Gamma3},\ref{jour}) are used for the calculation of the
singular part of the vacuum current in Appendix D.1.

\end{document}